\documentclass[twocolumn]{aastex631}
\usepackage{ulem}
\usepackage{mathrsfs}
\usepackage{amsmath}

\begin{document}

\newcommand{\msun}{{M_\odot}}
\newcommand{\rsun}{{R_\odot}}
\newcommand{\lsun}{{L_\odot}}
\newcommand{\tess}{{TESS}}
\newcommand{\gaia}{{\em Gaia}}
\newcommand{\rapid}{{{\tt rapid}}}
\newcommand{\delayed}{{{\tt delayed}}}
\newcommand{\porb}{{{P_{\rm{orb}}}}}
\newcommand{\metal}{{{\rm Z}}}

\newcommand{\mlc}{{{M_{\rm{LC}}}}}
\newcommand{\llc}{{{L_{\rm{LC}}}}}
\newcommand{\myr}{{{\rm Myr}}}
\newcommand{\mbh}{{{M_{\rm{BH}}}}}
\newcommand{\mns}{{{M_{\rm{NS}}}}}
\newcommand{\mwd}{{{M_{\rm{WD}}}}}
\newcommand{\mco}{{{M_{\rm{CO}}}}}

\newcommand{\mtot}{{M_{\rm{tot}}}}
\newcommand{\ecc}{{{Ecc}}}
\newcommand{\yr}{{{\rm{yr}}}}
\newcommand{\gyr}{{{\rm{Gyr}}}}

\newcommand{\kpc}{{{\rm{kpc}}}}
\newcommand{\pc}{{{\rm{pc}}}}
\newcommand{\BH}{{\rm{BH}}}
\newcommand{\LC}{{\rm{LC}}}
\newcommand{\cosmic}{{{\texttt{COSMIC}}}}
\newcommand{\mwdust}{{\texttt{mwdust}}}
\newcommand{\ccb}{\texttt{CCB22}}
\newcommand{\gaiamock}{{\texttt{gaiamock}}}
\newcommand{\astromet}{{\texttt{astromet}}}
\newcommand{\kira}{{\texttt{kira}}}
\newcommand{\sse}{{{\texttt{SSE}}}}
\newcommand{\bse}{{{\texttt{BSE}}}}
\newcommand{\Mmin}{{M_{\rm{min}}}}
\newcommand{\Mmax}{{M_{\rm{max}}}}
\newcommand{\sma}{{a}}
\newcommand{\fallback}{{\rm{FB}}}
\newcommand{\bhlc}{{\rm{BH\text{--}LC}}}
\newcommand{\nslc}{{\rm{NS\text{--}LC}}}
\newcommand{\nsms}{{\rm{NS\text{--}MS}}}
\newcommand{\nspms}{{\rm{NS\text{--}PMS}}}
\newcommand{\wdlc}{{\rm{WD\text{--}LC}}}
\newcommand{\colc}{{\rm{CO\text{--}LC}}}
\newcommand{\mtwelve}{{\bf{m12i}}}
\newcommand{\beam}{{{\rm{RB}}}}
\newcommand{\kms}{{\rm{km\ s^{-1}}}}
\newcommand{\tdur}{{t_{\rm{dur}}}}
\newcommand{\snr}{{{\rm{SNR}}}}
\newcommand{\days}{{\rm{day}}}
\newcommand{\ergspers}{{{\rm{ergs\ s^{-1}}}}}
\newcommand{\macc}{{M_{\rm{acc}}}}
\newcommand{\racc}{{R_{\rm{acc}}}}
\newcommand{\mzams}{{M_{\rm{ZAMS}}}}
\newcommand{\rzams}{{R_{\rm{ZAMS}}}}
\newcommand{\mch}{{M_{\rm{Ch}}}}
\newcommand{\qmin}{{q_\mathrm{min}}}
\newcommand{\qcrit}{{q_\mathrm{crit}}}
\newcommand{\nss}{{\em NSS}}
\newcommand{\mprim}{{m_{\rm{pri}}}}
\newcommand{\msec}{{m_{\rm{sec}}}}
\newcommand{\rpri}{{R_{\rm{pri}}}}
\newcommand{\rsec}{{R_{\rm{sec}}}}
\newcommand{\rRoche}{{R_{\rm{RL}}}}
\newcommand{\kmps}{{\rm{km\ s^{-1}}}}
\newcommand{\match}{{\rm{Match}}}
\newcommand{\nnslcmw}{{ N_{\nslc, \rm{MW} } }}
\newcommand{\nnslcsim}{{ N_{\nslc, \rm{sim} } }}

\newcommand{\sourav}[1]{\textbf{\color{red} Sourav: #1}}
\newcommand{\kb}[1]{\textbf{\color{teal} KB: #1}}
\newcommand{\cc}[1]{\textbf{\color{blue} CC: #1}}

\title{\gaia's promise to detect compact-object binaries II:\\
Predictions for dormant neutron star binaries in DR4 and beyond}
\author[0000-0001-9685-3777]{Chirag Chawla}
\affiliation{Astrophysics Research Center of the Open university (ARCO), The Open University of Israel, P.O Box 808, Ra’anana 4353701, Israel}
\email{chiragchawla134@gmail.com}
% \date{28th June 2021}

\author[0000-0002-3680-2684]{Sourav Chatterjee}
\affiliation{Department of Astronomy and Astrophysics, Tata Institute of Fundamental Research, Homi Bhabha Road, Navy Nagar, Colaba, Mumbai, 400005, India}
\email{souravchatterjee.tifr@gmail.com}

\author[0000-0001-5228-6598]{Katelyn Breivik}
\affiliation{Department of Physics, McWilliams Center for Cosmology and Astrophysics, Carnegie Mellon University, Pittsburgh, PA 15213, USA}

\begin{abstract}
\label{sec:abstract}
With the fourth data release (DR4) from \gaia\ imminent, we predict neutron star (NS) binaries with luminous companions (LCs) expected to be detectable in DR4 and the final data release (DR5). We create realistic synthetic populations of \nslc\ binaries in the Milky Way using \cosmic, assuming (a) observationally motivated initial distributions of binary properties, (b) metallicity and location dependent star formation history of the MW adopted from the FIRE-2 galaxy m12i, and (c) all relevant binary interaction physics, with the rapid and delayed supernova prescriptions. We estimate the {\em detectable} \nslc\ population using an approximate \gaia\ selection function based on projected orbital size of the LC as well as \gaiamock, which emulates \gaia's multi-epoch astrometry and orbit-fitting pipeline- both after taking into account extinction and reddening. Across supernova models and statistical fluctuations, the former (\gaiamock) pipeline predicts $\sim14$--$29$ ($\sim1$--$8$), $4500$--$4700$ ($100$--$110$), and $4800$--$5000$ ($300$--$380$) detectable \nslc\ binaries in DR3, DR4, and DR5, respectively. While the orbital-size based pipeline brackets the DR3 yield of \nslc\ candidates, the \gaiamock\ prediction falls short, suggesting uncertainties in the modeled binary evolution or contributions from other formation channels. DR4 is expected to directly shed light on the the relative importance of detectable NS-LC binaries from isolated binary evolution vis a vis other channels such as dynamical formation. Interestingly, our models suggest that future releases will also include $22^{+7}_{-5}\ (13^{+5}_{-4})$ and $124^{+14}_{-15}\ (53^{+10}_{-6})$ \gaiamock-detectable pulsar--MS binaries in DR4 and DR5, respectively, for the \rapid\ (\delayed) model, which are prime targets for radio follow-up.
\end{abstract}

\section{Introduction}
\label{S:intro}

Neutron stars (NSs) are among the densest objects in the Universe \citep[][]{Landau_1932}. Their formation is linked to the collapse of massive stars ($\gtrsim8\,\msun$) that explode as supernovae \citep[SNe;][]{Baade_1934a, Baade_1934b, Baade_1934c}, leaving behind a stable core supported by the degeneracy pressure of matter at nuclear density \citep[][]{Tolman_1934, Oppenheimer_1939}. NSs manifest as a rich variety of exotic astrophysical sources, including pulsars \citep[][]{Hewish_1968}, magnetars \citep[][]{Mazets_1979a, Mazets_1979b}, and likely fast radio bursts \citep[FRBs;][]{Lorimer_2007}. When paired with another compact object (CO) or a stellar companion, they give rise to X-ray binaries \citep[XRBs;][]{Giacconi_1962}, binary NS (BNS) mergers \citep[][]{GW_170817a}, NS--black hole (BH) mergers \citep[][]{Etienne_2007, Foucart_2008, Abbott_2021}, short gamma-ray bursts \citep[sGRBs;][]{Lee_2007}, and binary millisecond pulsars \citep[][]{Bisnovatyi_1974, Backer_1982}, observed in both Galactic and extragalactic environments. Together, these sources provide unique probes of NS formation --- the SN mechanism \citep[e.g.,][]{Janka_2012, Burrows_2021} and the natal kicks imparted at birth \citep[e.g.,][]{Hobbs2005, Burrows_2024, Popov_2025} --- and of the equation of state (EoS) of NS matter \citep[e.g.,][]{Lattimer_2016, Ozel_2016}. They may also constrain the binary interaction processes --- such as common envelope evolution \citep[][]{Ivanova2013}, mass transfer \citep[][]{Tauris_2006, Tauris_2023}, and tidal evolution \citep[][]{Zahn_1977, Hut_1981} --- that shape their present-day orbital, stellar, and kinematic properties.

Early NS discoveries include the isolated radio pulsar PSR~B1919+21 \citep[][]{Hewish_1968}, the binary pulsar PSR~B1913+16 \citep[][]{Hulse_1975}, and the bright X-ray source Scorpius~X-1 \citep[][]{Giacconi_1962}, which was later shown to be powered by an accreting NS \citep[][]{Cowley_1975, Fomalont_2001, Steeghs_2002}. Today, $\sim4500$ pulsars \citep[][]{Manchester_2005} and $\sim300$ NS--XRBs have been identified \citep[][]{Liu_2006, Liu_2007, Avakyan_2023, Fortin_2023, Fortin_2024, Neumann_2023}. Ground-based gravitational-wave (GW) observatories have since added another probe of the NS population: the mergers of binaries containing NSs. Through the second part of the fourth observing run \citep[O4b;][]{GWTC_5}, the LIGO--Virgo--KAGRA (LVK) observatories have detected two BNS mergers \citep[][]{GW_170817a, GW_190425} and four NS--BH mergers \citep[GW200105, GW200115, GW230518, and GW230529;][]{Abbott_2021, GW_230529, GWTC_4}, while NS--white dwarf (WD) mergers, whose GW emission peaks at millihertz--decihertz frequencies below the LVK band, have yet to be detected \citep[][]{Toonen_2018, Zenati_2019, Kang_2024}. The BNS merger GW170817 was also accompanied by an sGRB and a kilonova, providing the first direct evidence linking BNS mergers to sGRBs \citep[][]{GW_170817b, GW_170817c, GW_170817d}.

X-ray and GW observations detect only the subset of NS binaries in orbits tight enough for the NS to accrete and emit X-rays or for the binary to merge within a Hubble time \citep[][]{Eastep_2026}. Many of the remaining NSs reside in wide binaries with a luminous companion (LC) from which they accrete no material; unless the NS is an active pulsar, such systems are invisible to both electromagnetic (EM) and GW observatories \citep[e.g.,][]{El-badry_2024b, Chawla_2025}. These binaries can therefore be identified only through the orbital motion that the NS gravitationally induces in its LC \citep[][]{Harrison_1965, Kamp_1975}. \citet[][]{Zeldovich_1966} and \citet[][]{Trimble_1969} presented the first catalogs of $\sim50$ dormant \colc\ binary candidates, identified from the radial-velocity orbits of single-lined spectroscopic binaries. These catalogs also claimed the first dormant NS candidate, around $\alpha$~Her~B, which was later ruled out by high-resolution spectroscopic observations \citep[][]{Thiering_1993}. More recently, advances in spectroscopic and photometric sky surveys have enabled the discovery of several dormant \nslc\ candidates, e.g., 2XMM~J1255 \citep[][]{Mazeh_2022}, 2MASS~J06163552 \citep[][]{Yuan_2022}, LAMOST~J1123 \citep[][]{Yi_2022}, LAMOST~J2354 \citep[][]{Zheng_2023}, 56~UMa \citep[][]{Escorza_2023}, 2MASS~J1527 \citep[][]{Lin_2023, Zhang_2024}, J0606+2132 \citep[][]{Zhao_2024a}, G3431, G8441, and G4031 \citep[][]{Zhao_2024b}, and J0008+0128 \citep[][]{Zhao_2026}.

Several theoretical studies predict $\sim10^{8}$--$10^{9}$ NSs in the Milky Way \citep[MW;][]{Brown1994, Sartore_2010}. The vast majority of these are expected to be isolated \citep[][]{Popov_2000}, since only $\sim5$--$10\%$ of NSs form in the first SN of a binary and remain bound to their companion \citep[][]{Kochanek_2019, Renzo_2019, Chrimes_2022}. This is because, although most massive stars are expected to be born in binaries or higher-order multiples \citep[][]{Sana_2008, Sana_2009, Sana_2010, Sana_2011, Sana_2012, Moe_2017, Offner_2023}, most of these systems are destroyed by stellar mergers or by SN kicks of $\sim 250$\,km\,s$^{-1}$ \citep[][]{Hobbs2005, Faucher-Giguere_2006}. Accordingly, more than $99\%$ of young pulsars are isolated \citep[][]{Lorimer_2008}. Population synthesis models estimate that $\sim10^{4}$--$10^{5}$ NSs are in detached binaries with a non-degenerate stellar companion \citep[][]{Willems_2003, Afonina_2026, Chawla_2025}. Most recently, \citet[][]{Shariat_2026} estimate that only $0.01$--$1\%$ of all NSs are born into au-scale orbits with solar-type companions. Once selection effects due to the parallax, the intrinsic brightness of the LC, the orbital separation, and the coverage and precision of the surveys are taken into account, $\sim100$--$5000$ dormant \nslc\ binaries are expected to be discovered in the coming decade \citep[][]{Yalinewich2018, Wang_2022, Chawla_2025, Eastep_2026}. \gaia\ \citep[][]{Gaia_2016} promises to play a key role in discovering these dormant \nslc\ binaries through its decade-long astrometric, photometric, and spectroscopic measurements. This began with the DR3 non-single-star (\nss) catalog \citep[][]{Arenou_2023}, which demonstrated \gaia's capability by yielding a small population of 21 dormant \nslc\ candidates with spectroscopically validated orbits \citep[][]{Andrews_2022a, Shahaf_2022, El-badry_2024a, El-badry_2024b}, along with further candidates \citep[][]{El-badry_2026, Simon_2026}. In addition, the NS mass ($\mns$) distribution of these \gaia\ \nslc\ candidates closely resembles that of the first-born (recycled) NSs in Galactic double NS binaries \citep[][]{Schiebelbein-Zwack_2026b}.

In \citet[][]{Chawla_2025}, we provided a census of the dormant compact object-luminous companion (\colc) binaries expected over \gaia's end-of-mission (EOM) observing baseline, and showed that our modeled \nslc\ population is in close agreement with the \nslc\ candidates identified in DR3. This motivates us to extend our previous analysis to provide a more comprehensive picture of the expected \nslc\ population in the upcoming DR4 and final DR5 releases. The paper is structured as follows. Section~\ref{sec:numerical_setup} describes our modeling setup and the \gaia\ selection functions adopted for DR3, DR4, and DR5. In Section ~\ref{sec:results}, we present the key properties of the MW's \nslc\ population and of its \gaia-observable subset. Section~\ref{sec:atnf} compares the pulsars in our modeled \nsms\ binaries with known pulsar--main-sequence binaries in the ATNF pulsar catalog. We conclude in Section~\ref{sec:conclusions}.

\section{Numerical setup}
\label{sec:numerical_setup}

We use the Python-based, open-source rapid binary population synthesis (BPS) suite \cosmic\ \citep[][]{Breivik_2020, Breivik_2021} to generate realistic populations of \nslc\ binaries in the MW. \cosmic\ integrates the single stellar evolution (\sse) fitting formulae of \citet[][]{Hurley2000} with the binary stellar evolution (\bse) algorithm of \citet[][]{Hurley2002}, and includes revised modelling prescriptions for stellar evolution physics, binary interaction processes, stellar winds, and SN physics \citep[][]{Breivik_2020}.

\subsection{Initialised binary population}
\label{sec:initial_cond}
\cosmic\ begins by generating a zero-age main sequence (ZAMS) binary population, drawing the stellar and orbital parameters of each binary from observationally motivated probability distribution functions. We adopt the `independent' sampling scheme of \cosmic, which assumes no correlations at birth between the stellar parameters (e.g., primary mass and mass ratio) and the orbital parameters (e.g., semi-major axis and eccentricity). We sample the primary mass ($\mprim$) from the \citet[][]{Kroupa2001} initial mass function between $0.08$ and $150\,\msun$, but only evolve binaries with $3<\mprim/\msun<60$, since \cosmic\ assumes that only primaries in this range can form NSs depending on metallicity and binary evolution scenarios \citep[][]{Breivik_2020}. We assign the secondary mass ($\msec$) from a uniform mass-ratio ($q\equiv\msec/\mprim$) distribution \citep[][]{Mazeh1992, Goldberg1994, Kobulnicky_2007}, with $\msec\geq0.08\,\msun$. Initial eccentricities follow a thermal distribution \citep[][]{Jeans_1919, Ambartsumian_1937, Heggie1975}, and the initial semi-major axis ($\sma$) is drawn from a log-uniform distribution \citep[][]{Abt1983} between a lower bound set by $\rpri\le\rRoche/2$ \citep[][]{Dominik_2012} and an upper bound of $10^{5}\,\rsun$, where $\rpri$ and $\rRoche$ are the radius and Roche-lobe radius of the primary.

We use a metallicity-dependent star-formation history based on the final snapshot of the synthetically modeled galaxy {\bf m12i} from the Latte suite of FIRE-2 galaxies \citep[][]{Hopkins2018, Wetzel2016}. We choose {\bf m12i} because it closely resembles the MW in terms of stellar and gas mass, size, and stellar morphology \citep[][]{Wetzel2016, Sanderson2020}. From the {\bf m12i} star particles, we sample the initial metallicities ($\metal$) and ages of the ZAMS binaries. Since the \bse\ prescriptions are limited to $10^{-4}\le\metal\le0.03$, sampled metallicities outside this range are set to the nearest limiting value before being passed to \cosmic\ as input. We further use the \textsf{ananke} framework to assign three-dimensional Galactic positions to each binary by resampling from the smoothed density distribution of the star particles \citep[\autoref{sec:Synthetic MW};][]{Sanderson2020}. This approach preserves the complex correlations between age, metallicity, and the spatial distribution of the star particles in {\bf m12i}.

\subsection{Stellar and binary evolution model}
\label{sec:sse_bse_model}
\cosmic\ uses the fitting formulae of \sse\ \citep[][]{Hurley2000} to evolve the stellar properties of the binary components from the ZAMS to the present day. At each step, it employs the \bse\ algorithm \citep[][]{Hurley2002} to evolve the orbital-period ($\porb$), the eccentricity ($\ecc$), and the orbital angular momentum through binary interaction processes such as mass transfer (MT) through Roche-lobe overflow (RLOF), mass loss through stellar winds, and tidal evolution. Beyond single stellar evolution, these binary interaction phases may also affect the stellar properties, such as mass, radius, and luminosity, and hence the subsequent evolution of the binary components.

At the onset of RLOF of the donor, the stability of MT is determined using the radius--mass exponents, $\zeta$ \citep[][]{Webbink_1985}. In \bse/\cosmic, these radius--mass exponents are further translated into a simpler comparison between the mass ratio of the binary ($q=m_{\rm don}/m_{\rm acc}$) at the onset of RLOF and the critical mass ratio, $\qcrit$, derived assuming $\zeta_{\rm ad}=\zeta_{\rm L}$ \citep[][]{Tout1997}. We adopt the standard \bse\ $\qcrit$ values from \citet[][]{Hurley2002} along with those of \citet[][]{Hjellming_1987} for giant branch (GB) and asymptotic giant branch (AGB) stars.

When $q<\qcrit$, MT is dynamically stable and, depending on the evolutionary stage of the donor, proceeds on its nuclear or thermal timescale, at the rates given by \citet[][]{Tout1997} and \citet[][]{Hurley2002}. When $q>\qcrit$, MT proceeds on a dynamical timescale and leads to a common-envelope \citep[CE;][]{Paczynski1976} phase if the donor is a giant-like star with a distinct core--envelope structure, or to a merger otherwise \citep[][]{Tout1997, Hurley2002}. We model CE evolution (CEE) with the energy formalism, parameterized by $\alpha_{\rm CE}$ \citep[][]{Livio_1984} and $\lambda$ \citep[][]{deKool_1990}. The parameter $\alpha_{\rm CE}$ is the fraction of the orbital energy released during CEE that is used to unbind the envelope, while $\lambda$ parameterizes the binding energy of the envelope, which can include contributions from its thermal and recombination energies in addition to its gravitational binding energy \citep[][]{Han_1995}. Following the standard model of \citet[][]{Claeys2014}, we adopt $\alpha_{\rm CE}=1$ \citep[][]{Regos_1995} and their variable $\lambda$, which depends on the type, mass, and luminosity of the donor (their Appendix A) and includes the thermal energy of the envelope but not its recombination energy.

\subsection{SN models: NS formation and kicks}
\label{sec:SN_models}

We adopt two widely used models for NS formation through core-collapse supernovae (CCSNe), \rapid\ and \delayed\ \citep[][]{Belczynski_2012, Fryer2012}. The two models differ in the time between core bounce and the explosion, which is set by the timescales on which neutrino-driven instabilities are allowed to grow \citep[][]{Colgate_1966, Boccioli_2025}. In the \delayed\ model, the longer time before the explosion allows the proto-NS to accrete more mass, and the weaker explosion leads to more fallback, producing more massive NSs ($\Delta\mns\sim1\,\msun$). In both models, we adopt a maximum NS mass of $3\,\msun$, above which compact remnants are treated as BHs \citep[][]{Chawla2022, Chawla_2023, Chawla_2025}.

We also include NS formation through electron-capture collapse. Stars with $8\lesssim M_{\rm ZAMS}/\msun\lesssim10$ burn carbon after helium but stop short of neon, leaving an oxygen-neon (ONe) core whose mass determines their fate. If this core mass exceeds the Chandrasekhar mass, $M_{\rm Ch}\approx1.38\,\msun$ \citep[][]{Chandrasekhar_1931}, electron captures onto Mg and Ne nuclei initiate its collapse to a low-mass NS in an electron-capture SN \citep[ECSN; e.g.,][]{Miyaji_1980, Nomoto_1984, Nomoto_1987}; otherwise, an ONe WD remains.\footnote{The $M_{\rm ZAMS}$ ranges are indicative and the nature of the remnant depends on the pre-collapse core mass. As a result, metallicity-dependent stellar winds and binary interactions can significantly blur these boundaries \citep[][]{Mink_2009, Langer_2012, Agrawal_2022, Eldridge_2022}.} In our models, ECSNe occur for He-core masses of $1.4$--$2.5\,\msun$, which produce ONe cores that reach $M_{\rm Ch}$ \citep[][]{Podsiadlowski_2004}. An ONe WD can also collapse to an NS through accretion-induced collapse (AIC) when it accretes from its companion via RLOF until its mass exceeds $M_{\rm Ch}$ \citep[][]{Nomoto_1991, Hurley2002}.

Owing to deviations from spherical symmetry during the collapse and subsequent explosion, NSs may receive a natal kick \citep[][]{Fryer_2001, Janka_2013}. We draw NS kick magnitudes from a bimodal Maxwellian distribution with $\sigma=265\,{\rm km\,s^{-1}}$ for CCSNe \citep[][]{Hobbs2005} and $\sigma=20\,{\rm km\,s^{-1}}$ for ECSNe and AICs \citep[][]{Podsiadlowski_2004}.

\subsection{Synthetic galactic population}
\label{sec:Synthetic MW}
\cosmic\ iteratively simulates binaries until the distributions of observable parameters including $\mns$, the LC mass ($\mlc$), $\porb$, and $\ecc$ converges to a high degree of resolution. We quantify the convergence of the distributions using a `$\match$' parameter inspired by the matched filtering technique \citep[][]{Chatziioannou2017, Breivik_2020}. For this study we have adopted $\log_{10}(1-\match)=-5$. The matched-filtering technique enables \cosmic\ to evolve only a fraction of the MW mass to generate a population of \nslc\ binaries, referred to as the intrinsic population, which accurately captures the parameter space of the full \nslc\ binary population in the MW. To estimate the total number of \nslc\ binaries in the MW, we scale the intrinsic \nslc\ population by a factor equal to the ratio of the MW mass ($M_{\rm{\bf{m12i}}}$ as specified by {\bf m12i}) to the total simulated mass ($M_{\rm{sim}}$ including both single and binary stars determined based on the adopted binary fraction) which created the converged intrinsic \nslc\ population. The number of \nslc\ binaries in the MW ($\nnslcmw$) can then be estimated from the number of simulated \nslc\ binaries ($\nnslcsim$) as-

\begin{equation}
   \nnslcmw = \nnslcsim \frac{M_{\rm{\bf m12i}}}{M_{\rm{sim}}}.
\end{equation}

We sample $N_{\nslc,\rm{MW}}$ binaries with replacement from the intrinsic population of \nslc\ binaries assigning all the key stellar and orbital properties. Each binary is also given a unique position in the MW galaxy based on the age and metallicity of the ZAMS progenitor. This preserves the complex correlations between the position, age, and metallicity of the different star particles in ${\bf m12i}$. Every binary is also given a random orientation with respect to the line of sight. Based on the orientation of the binary, we calculate its sky-projected orbit using the Thiele-Innes elements \citep[see][for a detailed description]{Chawla2022}. Finally, we create multiple Galactic realisations of the \nslc\ binary populations in order to incorporate the effects of statistical fluctuations due to distribution of the binaries in the MW and their orbital orientations. We create 100 Galactic realisations for the \nslc\ binary population. 

We evaluate the \gaia-relevant magnitudes of the LCs from the bolometric luminosities and effective temperatures given by \cosmic\ using the {\tt isochrones} package \citep[][]{Isochrones_2015}. We correct these magnitudes to incorporate the effect of reddening and extinction due to interstellar dust using the three-dimensional `combined19' dust map included in the $\mwdust$ package \citep[][]{Drimmel2003, Marshall2006, Green2019, Bovy2016} given the distance and Galactic location of each binary.

\subsection{\gaia\ selection function (GSF)}
\label{sec:gaia_sel_function}
For our synthetic Galactic population of \nslc\ binaries, we first use the approximate \gaia\ selection function of \citet[][]{Chawla2022}, hereafter the \ccb\ criteria, to identify the \gaia-detectable subset. These criteria are based on the apparent $G$ magnitude of the LC and the astrometric motion induced by its orbit around the NS. As a more realistic alternative, we assess detectability with \gaiamock\ \citep[][]{El-badry_2024d}, which emulates the \gaia\ NSS pipeline.

\subsubsection{GSF: \ccb\ criteria}
\label{sec:GSF-chawla}
We use the \ccb\ criteria, based on the projected orbital motion of the LC \citep[][]{Chawla2022}, to test the detectability of an \nslc\ binary by \gaia. For detection in DR5, the \ccb\ criteria require
\begin{enumerate}
    \item $\porb/\yr<10$,
    \item $G<20$,
    \item $\alpha_{\rm LC}>\sigma_{G}$,
\end{enumerate}
where $\alpha_{\rm LC}$ is the sky-projected angular size of the LC's orbit and $\sigma_{G}$ is the single-pointing astrometric precision of \gaia, adopted from \citet[][]{Lindegren_2021}. For DR4, we further require the DR5-detectable binaries to have
\begin{enumerate}\setcounter{enumi}{3}
    \item $\porb/\yr \leq 5.5$.
\end{enumerate}
Finally, for DR3, we apply additional selection cuts from the DR3 NSS pipeline \citep[][]{Arenou_2023, Halbwachs_2023} to the DR4-detectable population:
\begin{enumerate}\setcounter{enumi}{4}
    \item $G \leq 19$,
    \item $\porb/\yr \leq 3$, 
    \item $\Pi/\sigma_{\Pi} \geq 20{,}000\,{\rm d}/\porb$,
\end{enumerate}
where $\Pi$ and $\sigma_{\Pi}$ denote the parallax and its uncertainty, respectively.

\subsubsection{GSF: \gaiamock}
\label{sec:GSF-gaiamock} Following recent \gaia\ data releases, several synthetic \gaia\ selection functions have been developed, including \astromet\ \citep[][]{Penoyre_2022, Penoyre_2022a}, the selection function of \citet[][]{Lam_2025}, and \gaiamock\ \citep[][]{El-badry_2024d}. We adopt \gaiamock\ because it emulates the astrometric model-fitting cascade of the \gaia\ NSS pipeline \citep[][]{Arenou_2023, Holl_2023a}. It has been shown to reproduce the DR3 NSS catalog itself \citep[][]{El-badry_2024d}, as well as the \bhlc\
\citep[][]{Nagarajan_2025a} and \wdlc\ \citep[][]{El-badry_2024d, Yamaguchi_2025} samples identified in it \citep[][]{El-badry2022e, El-badry_2023, Shahaf_2024}.

We apply \gaiamock\ to the \nslc\ binaries that satisfy cuts 1--3 of the \ccb\ selection function (i.e., the DR5-detectable population; \autoref{sec:GSF-chawla}) to test how many of these binaries would receive an orbital solution in \gaia\ DR3, DR4, and DR5. For each binary, we simulate the \gaia\ epoch astrometry using its sky position, parallax, orbital parameters, and orientation angles from our synthetic population, together with its apparent $G$ magnitude including dust extinction. We treat the NS as a dark companion (i.e., a flux ratio of zero) and draw the time of periastron passage uniformly within one orbital period. We set the input proper motions to zero; since the proper motion is fitted as a free parameter in the cascade, this choice does not affect the resulting orbital solutions. We then fit the simulated astrometry with the full \gaiamock\ cascade, including the acceleration solutions. Only sources whose single-star solutions exceed a threshold renormalized unit weight error (RUWE) are passed to the binary models of the cascade. For this threshold, we adopt the values derived by \citet[][]{Guerriero_2026}, above which fewer than one in $10^6$ well-behaved single-star solutions are expected, i.e., RUWE $>1.25$, $1.15$, and $1.11$ for DR3, DR4, and DR5, respectively.

For detection in DR5, we require a binary to receive an orbital solution from the \gaiamock\ cascade. For DR4, following \citet[][]{Nagarajan_2025a}, we further require the orbital solution to have
\begin{enumerate}
    \item $a_0/\sigma_{a_0}>5$, 
    \item $\Pi/\sigma_{\Pi}>5$,
\end{enumerate}
where $a_0$ and $\sigma_{a_0}$ denote the photocentre semi-major axis and its uncertainty, respectively. Finally, for DR3, we instead apply the post-processing filters used to construct the DR3 NSS orbital catalog \citep[][]{Halbwachs_2023} to the resulting orbital solutions:
\begin{enumerate}\setcounter{enumi}{2}
    \item $a_0/\sigma_{a_0}>\max\left(5,\,158/\sqrt{\porb/{\rm d}}\right)$,
    \item $\Pi/\sigma_{\Pi}>20{,}000\,{\rm d}/\porb$, 
    \item $\sigma_e<0.079\ln(\porb/{\rm d})-0.244$,
\end{enumerate}
where $\sigma_e$ denotes the eccentricity uncertainty.

\section{Results}
\label{sec:results}
\begin{deluxetable*}{c|c|c|c|c|c|c|c|c|c}
\tablecolumns{10}
\tabletypesize{\normalsize}
\tablecaption{\nslc\ binaries in the Milky Way}
\tablehead{
 \colhead{SN Model}\vline &
 \colhead{LC Type}\vline & \multicolumn{2}{c}{Intrinsic }\vline &
 \multicolumn{6}{c}{\gaia\ Resolvable} \\
 \cline{1-10}
\colhead{}\vline &
 \colhead{}\vline & \colhead{all detached}\vline &
 \colhead{$\porb/\yr\le 10$}\vline &
 \multicolumn{3}{c}{\ccb}\vline &
 \multicolumn{3}{c}{\gaiamock}\\
 \cline{1-10}
 \colhead{}\vline &
 \colhead{}\vline & \colhead{}\vline & \colhead{}\vline &
 \colhead{dr5 }\vline & \colhead{dr4 }\vline & \colhead{dr3}\vline &
 \colhead{dr5 }\vline & \colhead{dr4 }\vline & \colhead{dr3}
 }
\startdata
%  %
  & MS & $103,938^{+233}_{-219}$ & $98,199^{+245}_{-199}$ &
  $4553^{+73}_{-87}$ & $4279^{+66}_{-86}$ & $27^{+7}_{-6}$ &
  $335^{+24}_{-21}$ & $87^{+11}_{-10}$ & $6^{+3}_{-3}$\\
  \rapid & PMS & $19,942^{+111}_{-115}$ & $19,465^{+126}_{-110}$ &
  $290^{+27}_{-25}$ & $252^{+23}_{-23}$ & $2^{+3}_{-1}$ &
  $44^{+8}_{-9}$ & $17^{+6}_{-5}$ & $2^{+2}_{-1}$\\
  & Total & $123,895^{+246}_{-302}$ & $117,700^{+230}_{-247}$ &
  $4843^{+71}_{-84}$ & $4536^{+59}_{-90}$ & $29^{+9}_{-5}$ &
  $378^{+24}_{-22}$ & $103^{+13}_{-10}$ & $8^{+3}_{-3}$ \\
%   %
   \hline
%   %
  & MS & $108,409^{+226}_{-218}$ & $102,101^{+224}_{-272}$ &
  $4695^{+78}_{-84}$ & $4418^{+89}_{-80}$ & $14^{+5}_{-6}$ &
  $270^{+23}_{-17}$ & $91^{+10}_{-11}$ & $1^{+1}_{-1}$ \\
  \delayed\ & PMS & $20,574^{+111}_{-93}$ & $20,100^{+97}_{-100}$ &
  $334^{+21}_{-25}$ & $312^{+18}_{-23}$ & $0$ &
  $29^{+7}_{-6}$ & $17^{+7}_{-4}$ & $0$  \\
  & Total & $128,964^{+279}_{-195}$ & $122,170^{+269}_{-242}$ &
  $5021^{+89}_{-74}$ & $4738^{+80}_{-80}$ & $14^{+5}_{-6}$ &
  $301^{+20}_{-17}$ & $108^{+13}_{-10}$ & $1^{+1}_{-1}$\\
  %  %
    % \hline
\enddata
\tablecomments{\nslc\ numbers in the Milky Way predicted in our models. The numbers and errors denote the median and the spread between the $10$th and $90$th percentiles across the Milky-Way realisations ``Intrinsic" denotes the present-day entire detached population and the
detached \nslc\ binaries with $\porb/\yr\le10$  in the Milky Way. ``Resolvable" denotes the subset of \nslc\ binaries resolvable by \gaia's astrometry based on our earlier \citet[][]{Chawla2022}
\gaia\ selection function and \gaiamock\ from \citet[][]{El-badry_2024d}.
}
\label{tab:detection}
\end{deluxetable*}

In this section, we describe the key features of the present-day intrinsic \nslc\ populations in the MW, i.e., before any observational selection cuts are applied. We focus on the properties that \gaia\ can directly constrain or that strongly affect the detectability of these binaries: $\porb$, $\mns$, $\mlc$, and $\ecc$. We also highlight the relative importance of different binary interaction processes, such as MT and CE, and the impact of the two SN models on the distributions of these properties. These intrinsic distributions provide the baseline for understanding how the observational selection cuts shape the detectable \nslc\ population. Note that the short-period ($\porb/\yr\le10$) detached \nslc\ binaries make up nearly the entire detached population (\autoref{tab:detection}), and thus we restrict the discussion in the following subsections to this subset unless otherwise stated.

\subsection{Intrinsic \nslc\ Population}
\label{sec:intrinsic-prop}

\begin{figure*}
    \plotone{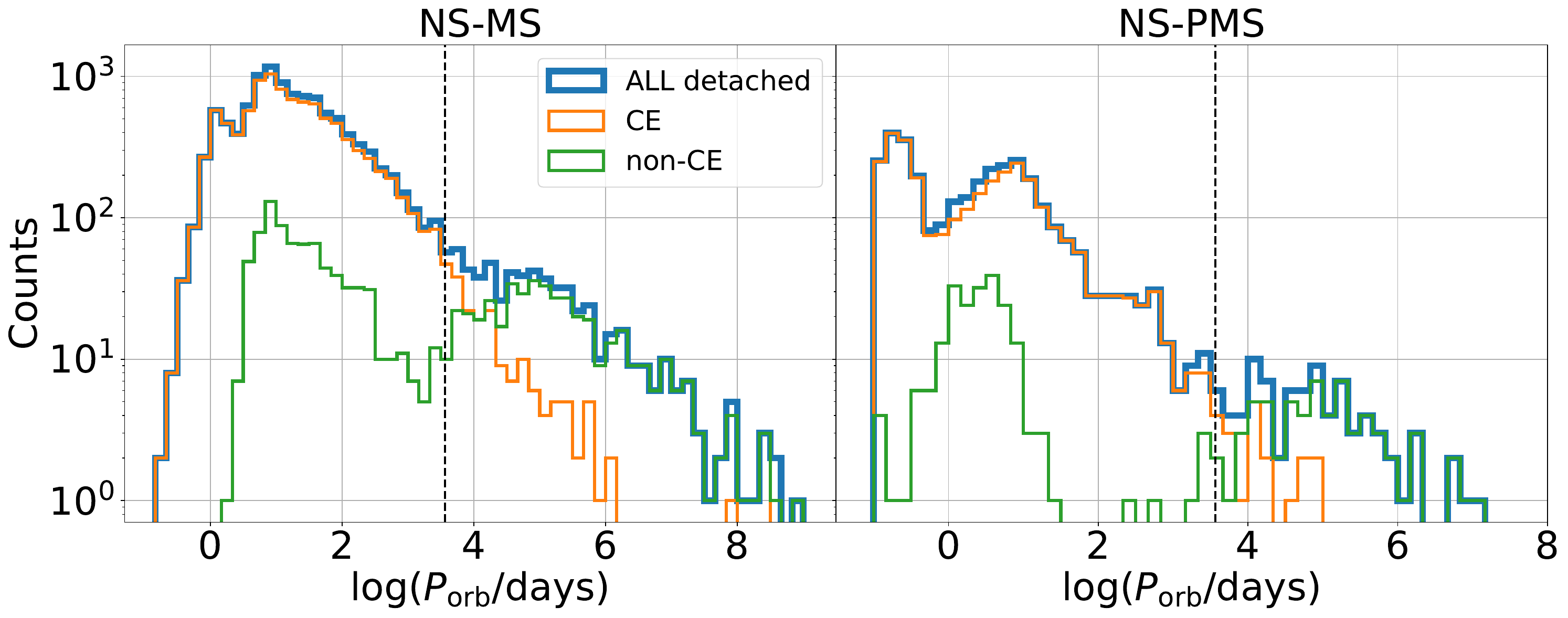}
    \caption{$\porb$ distribution of the present-day detached \nslc\ binaries for the \rapid\ model. Blue, orange, and green curves represent all detached \nslc\ binaries, those that went through at least one CE (initiated by the NS progenitor), and those that never went through a CE, respectively. The top and bottom panels represent NSs with an MS and a PMS companion, respectively. The dashed vertical line marks $\porb=10\,\yr$, the approximate duration of the \gaia\ mission. The corresponding figure for the \delayed\ model is shown in \autoref{fig:intrinsic-porb-delayed} in the Appendix.}
    \label{fig:intrinsic-porb}
\end{figure*}

\begin{figure}
    \plotone{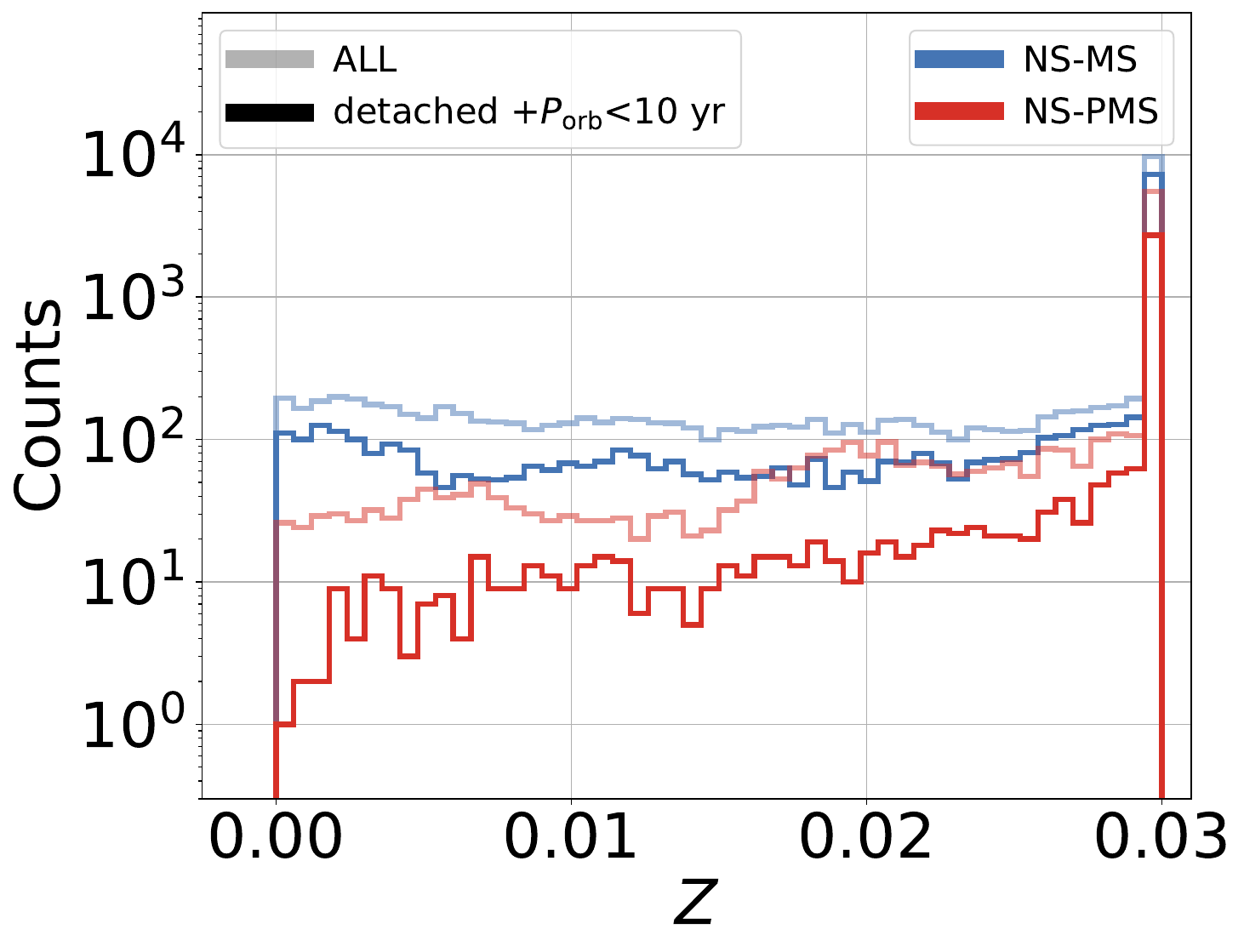}
    \caption{$\metal$ distribution of the present-day \nslc\ binaries for the \rapid\ model. Faded and bright curves represent all \nslc\ binaries and those that are detached with $\porb/\yr\le10$, respectively. Blue and red curves represent NSs with an MS and a PMS companion, respectively. The corresponding figure for the \delayed\ model is shown in \autoref{fig:intrinsic-met-delayed} in the Appendix.}
    \label{fig:intrinsic-met}
\end{figure}

\autoref{fig:intrinsic-porb} shows the present-day $\porb$ distribution of the detached \nslc\ binaries in the MW for the \rapid\ model; the distribution for the \delayed\ SN model is very similar. Unlike that of the \bhlc\ population \citep[][]{Chawla2022}, the $\porb$ distribution of the modeled \nslc\ population is not bimodal. This is a consequence of the stronger natal kicks and lower masses of NSs compared with BHs, which lead to the disruption of a large fraction of \nslc\ binaries in wider orbits. The present-day surviving \nslc\ population, including binaries that are currently transferring mass, also contains a much larger fraction of binaries that have experienced a CE during their evolution ($\sim94\%$) than the \bhlc\ population \citep[$\sim4$--$44\%$;][]{Chawla2022}. This is because the orbital shrinkage during the CE phase makes these \nslc\ binaries more likely to survive the relatively stronger natal kicks. Owing to the stronger kicks, the CE population of \nslc\ binaries extends to longer $\porb$ than that of \bhlc\ binaries. However, as with the \bhlc\ population, the \gaia-detectable \nslc\ population will include both binaries that have experienced a CE episode and those that have not. Among short-period ($\porb/\yr\le10$) \nslc\ binaries, $\sim93\%$ have experienced a CE, $\sim7\%$ have experienced stable MT only, and fewer than $0.1\%$ have never undergone MT during their evolution.

The metallicities (Z) of the \nslc\ binaries span $10^{-4}\lesssim\metal\lesssim0.03$ (\autoref{fig:intrinsic-met}). The metallicity distributions of both the \nsms\ and \nspms\ populations remain uniform across this entire range.\footnote{The apparent pile-up at $\metal\approx0.03$ is not physical: the BSE prescriptions are limited to $\metal\le0.03$, so \cosmic\ assigns $\metal=0.03$ to all binaries with higher metallicities (see \autoref{sec:initial_cond}).} This contrasts with the \bhlc\ population, where the abundance decreases with increasing metallicity \citep[][]{Chawla_2025} because higher metallicities increase the stellar wind mass-loss rates, which inhibit core growth and result in fewer BHs forming at high metallicity \citep[][]{Belczynski_2010, Romagnolo_2024, Romagnolo_2026, Chen_2026, Maclean_2026}. The resulting lower core masses also lead to stronger BH natal kicks, which can unbind more BH binaries \citep[][]{vanSon_2025, Chen_2026}. 
Interestingly, a similar lack of dependence between abundance and metallicity was found also for BNS mergers \citep[][and references therein]{vanSon_2025}. This suggests that, in these simulations, NS binaries remain nearly metallicity independent from the formation of the first NS until the BNS merger. In contrast, detailed BPS models, which self-consistently follow the response of stars to MT, find that the BNS merger efficiency increases toward low metallicity \citep[][]{Chattaraj_2026}. Rapid BPS codes such as \cosmic\ may therefore miss a metallicity dependence of NS binaries, particularly at low metallicity. More generally, since BPS outcomes can differ substantially between codes and physical assumptions, even for identical initial binaries \citep[][]{Belczynski_2022, Guerrero_2026}, the metallicity dependence of NS binary populations remains uncertain. Future observations of BNS mergers and \nslc\ binaries in significant numbers may shed light on this.

\begin{figure*}
    \plotone{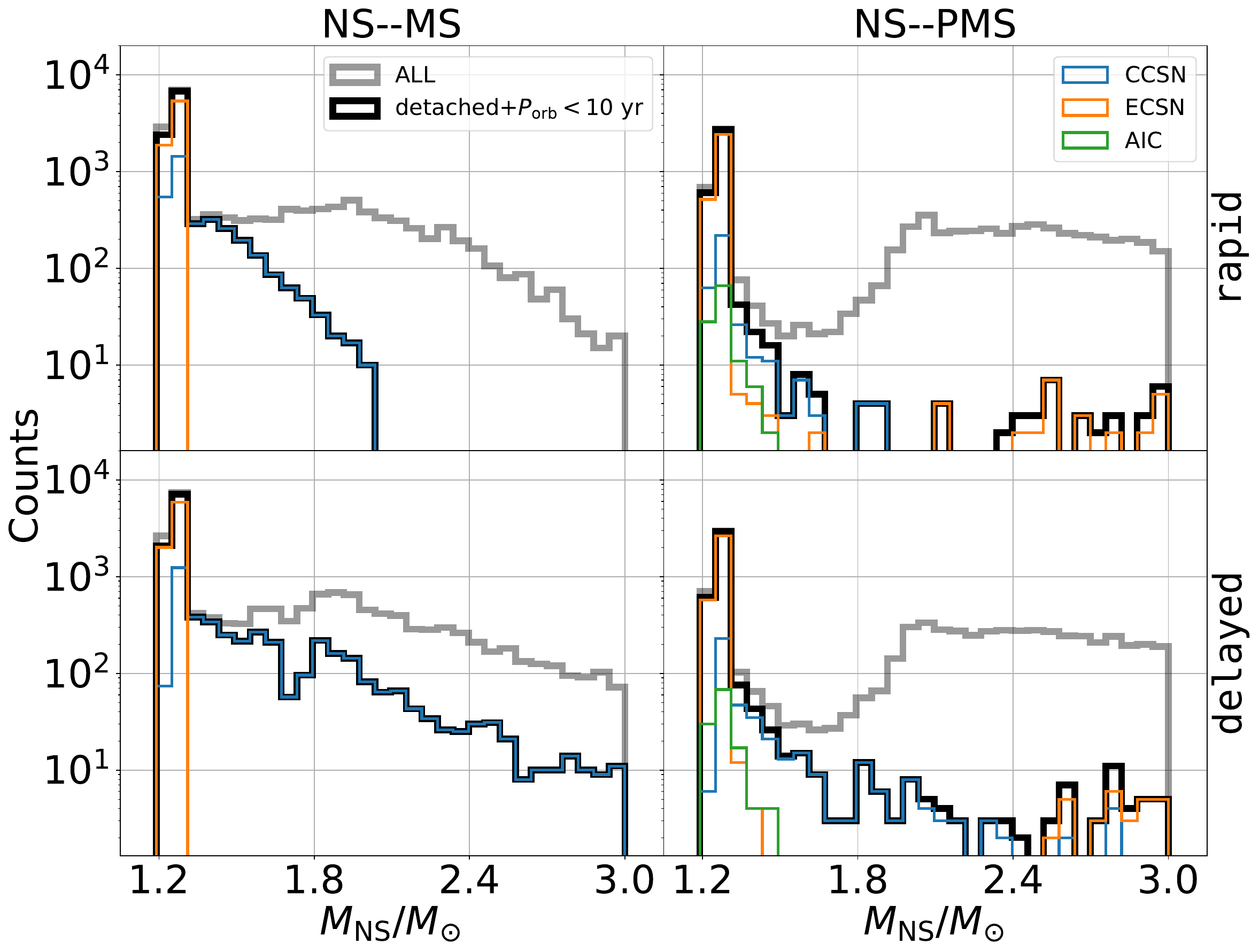}
    \caption{$\mns$ distribution of the present-day \nslc\ binaries.
    Gray and black curves represent all \nslc\ binaries and those that are detached with $\porb/\yr\le10$, respectively. Blue, orange, and green curves represent the subsets of the latter in which the NS formed through CCSN, ECSN, and AIC, respectively. The left and right columns represent NSs with an MS and a PMS companion, respectively,while the top and bottom rows represent the \nslc\ populations from the \rapid\ and \delayed\ models, respectively.}
    \label{fig:intrinsic-mns}
\end{figure*}

In \autoref{fig:intrinsic-mns}, we show the $\mns$ distributions of the \nslc\ binary population from both the \rapid\ and \delayed\ models. In case of BHs, the two models produce BHs with significantly different masses- most prominently, the \rapid\ model does not produce $\mbh$ between $3$ and $5\,\msun$ from CCSNe, but no such gap exists in the \delayed\ model \citep{Fryer2012, Belczynski_2012}. This gap also manifests in the \bhlc\ populations, both intrinsic and detectable, produced using the two models \citep{Chawla2022,Chawla_2023, Chawla_2025}. In contrast, these two SN models produce similar $\mns$ distributions in the full adopted range between $1.2$ and $3\,\msun$. This is clearly visible in the $\mns$ distribution for {\em all} \nslc\ binaries in our models. Despite this, the $\mns$ distributions for the detached short-period ($\porb/\yr\leq10$) \nslc\ populations produced in the two models differ significantly- while the \rapid\ model exhibit a declining $\mns$ distribution with $\mns/\msun\lesssim2.1$, $\mns$ extends all the way up to the adopted limit of $3\,\msun$ in the \delayed\ model. This difference does not arise from the cut on $\porb$, but from the restriction to detached binaries.

In the \rapid\ model, the explosion leads to a sharp transition between NSs and more massive BHs, with very few remnants in the $2$--$5\,\msun$ gap region ($18\%$ of remnants above $2\,\msun$ at solar metallicity and $<1.5\%$ at $0.1$ solar; \citealt{Fryer2012}). NSs with $\mns/\msun>2.1$ therefore form only through accretion from their LCs, and are found only in non-detached binaries. The absence of such NSs from the detached \rapid\ population is thus the lower end of the same mass gap seen for BHs. In the \delayed\ model, in contrast, far more remnants lie in the gap region ($33\%$ at solar metallicity and $17\%$ at $0.1$ solar metallicity; \citealt{Fryer2012}), so that such massive NSs also form directly in the SN. The proto-NS is more massive ($1.2$--$1.6\,\msun$), owing to the larger mass accreted during the convection phase \citep[][]{Fryer2012}. In addition, owing to the longer delay between core bounce and the explosion, the delayed engine has a lower explosion energy, which increases the amount of fallback and produces more massive NSs in the \delayed\ population. Since the fallback mass increases smoothly with the carbon--oxygen core mass, the remnant masses are continuous across the NS--BH boundary, and NSs with $\mns/\msun>2.1$ are also present in detached binaries. In the full \nslc\ population, the two $\mns$ distributions nevertheless appear similar, because accretion in non-detached binaries populates the high-mass end in the \rapid\ model. Note also that binary interactions such as MT and tides can reshape the pre-SN core of the NS progenitor, and hence $\mns$ \citep[][]{Mink_2009, Mink_2016, Song_2013, Song_2016, Marchant_2016, Laplace_2021, Schneider_2021, Schneider_2023}, changes that the versions of \sse\ and \bse\ used here do not follow and that we leave for future work. 

\begin{figure*}
    \plotone{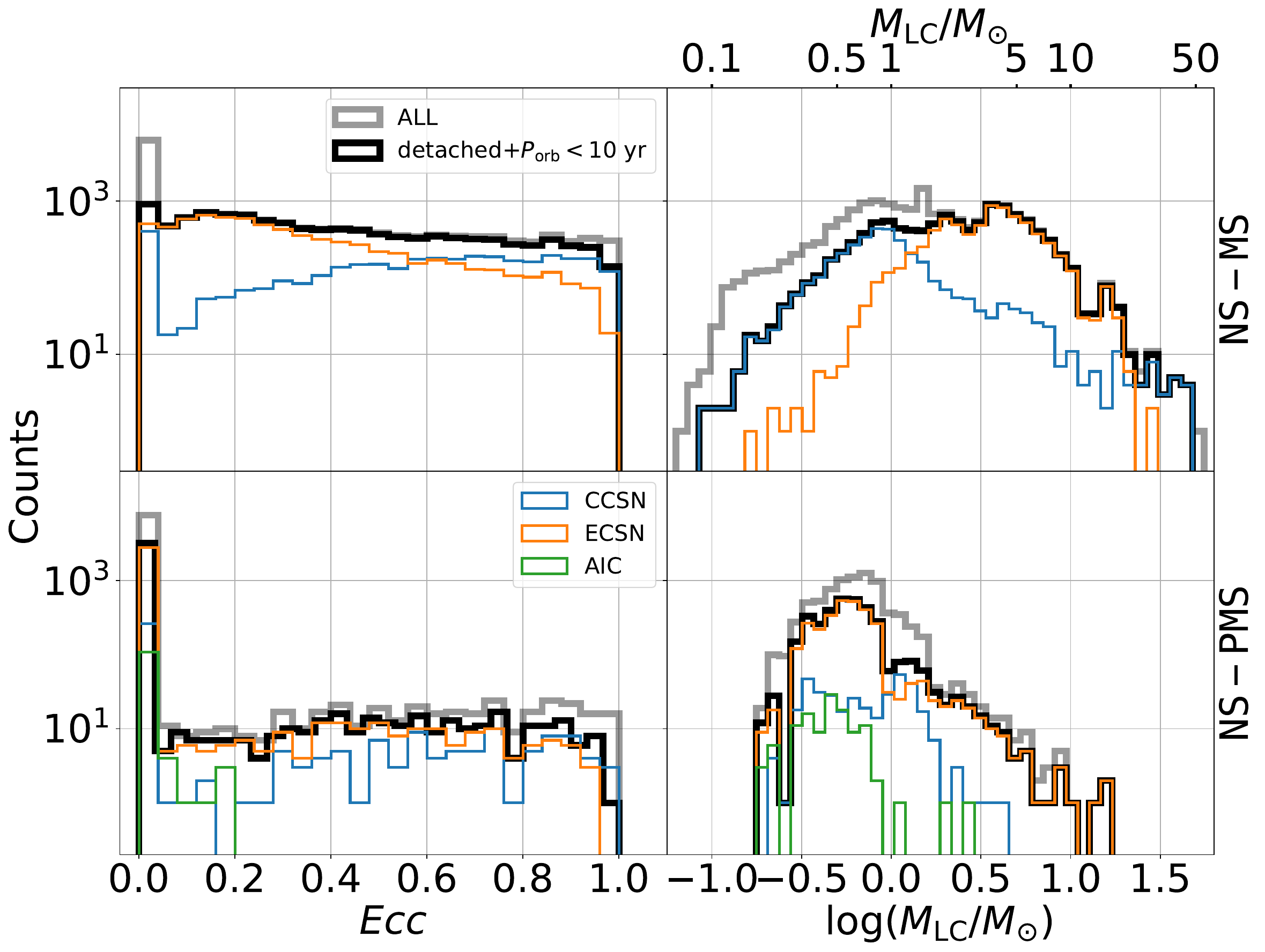}
    \caption{Same as \autoref{fig:intrinsic-mns}, but for the $\ecc$ (left) and $\mlc$ (right) distributions of the present-day \nslc\ binaries for the \rapid\ SN model. The top and bottom rows represent NSs with an MS and a PMS companion, respectively. The corresponding figure for the \delayed\ SN model is shown in \autoref{fig:intrinsic-ecc-mlc-delayed} in the Appendix.}
    \label{fig:intrinsic-ecc-mlc}
\end{figure*}

The $\ecc$ distributions of the \nsms\ binaries are mostly flat in both the \rapid\ and \delayed\ SN models (\autoref{fig:intrinsic-ecc-mlc}). About $3.5\%$ of the \nsms\ binaries have circular orbits ($\ecc=0$), and $\sim12\%$ have near-circular orbits ($0<\ecc\le0.1$). In contrast, most \nspms\ binaries ($\sim91$--$92\%$) are on circular orbits, because a large fraction ($\sim97\%$) of these circular \nspms\ binaries experienced a second MT/CE episode, initiated by the secondary, after NS formation. Eccentric ($\ecc\ge0.1$) binaries with NSs formed via ECSN, CCSN, and AIC account for only $5\%$, $3\%$, and $<1\%$ of the \nspms\ binaries, respectively. Both the ECSN and CCSN subsets of the \nsms\ binaries cover the entire $\ecc$ range between $0$ and $1$. Although the overall $\ecc$ distribution of the \nsms\ binaries appears flat, the two subsets follow opposite trends. The ECSN distribution peaks between $\ecc=0$ and $0.2$ and declines gradually toward $\ecc=1$, whereas the CCSN distribution falls sharply after a peak around $\ecc=0$ and then increases toward $\ecc=1$. This is again a consequence of the larger kick magnitudes for CCSN ($\sigma=265\kmps$) than for ECSN ($\sigma=20\kmps$). Since most \nslc\ binaries go through a CE phase before the primary collapses to form the NS, and the CE erases the memory of their initial eccentricity, their present-day eccentricities are set by the SN kicks and, for \nspms\ binaries, by subsequent circularisation through post-SN MT/CE episodes and tides.

The $\mlc$ distributions of the detached short-period \nslc\ binaries span nearly the same range as those of the entire population (including both detached and semi-detached binaries), although at $\mlc\lesssim1\,\msun$ the entire \nsms\ population is $\sim2.5$ times larger than its detached short-period subset. Similar to the $\mns$ distribution, the $\mlc$ distributions of the \nsms\ and \nspms\ populations are also distinct. The \nsms\ binaries span $\mlc\approx0.08$--$45\,\msun$, whereas the \nspms\ distribution is narrower, spanning $0.2$--$17\,\msun$. In both SN models, the $\mlc$ distribution of the \nsms\ population is multimodal. Separating the \nsms\ population by NS formation channel reveals two distinct distributions for the CCSN and ECSN subsets. The first peak, at $\mlc\sim0.85\,\msun$, arises from the CCSN subset. This is counterintuitive, since binaries with massive LCs are expected to be more likely to survive the larger CCSN kicks. However, this feature results from the combination of two effects. Since most \nsms\ binaries experience a CE phase, those with massive LCs emerge from the CE in relatively wide orbits, because a more massive companion requires less orbital shrinkage to eject the envelope. These wider post-CE orbits are then more easily disrupted by the large CCSN kicks. The second peak, at $\mlc\sim4\,\msun$, arises from the ECSN subset, which dominates above $\mlc\approx1.2\,\msun$. Most of these binaries ($\approx93\%$) also experienced a CE and, owing to their massive LCs, emerged from it in relatively wide orbits (median post-CE separation of $\approx27\,R_\odot$, compared with $\approx11\,R_\odot$ for CCSN-formed binaries with $\mlc\le1.2\,\msun$). These wide orbits survive the SN because ECSN kicks are much weaker (median $\approx30\,{\rm km\,s^{-1}}$), which is why this peak lies at larger $\mlc$. For the same reason, the CE-formed ECSN binaries extend to higher $\mlc$ than the CCSN-formed ones, outnumbering them by a factor of $\sim20$ above $\mlc=6\,\msun$.

Binaries with $\mlc>10\,\msun$ ($\approx3\%$ of the \nsms\ population) instead formed mainly through stable MT ($\approx80\%$ never experienced a CE), during which the LC accreted a median of $\approx8\,\msun$ from the NS progenitor, roughly doubling its initial mass, and was rejuvenated, which extends its remaining MS lifetime \citep[][]{Neo_1977, Hellings_1983, Dray_2007}. Among these, LCs of $10$--$20\,\msun$ are mostly paired with ECSN-formed NSs, whereas LCs above $20\,\msun$ are mostly paired with CCSN-formed NSs. This is because the CCSN progenitors in these binaries are more massive (median ZAMS mass of $\approx28\,\msun$, compared with $\approx12\,\msun$ for the ECSN progenitors) and therefore transfer more mass to the LC (median of $\approx15\,\msun$, compared with $\approx8\,\msun$). LCs above $10\,\msun$ are rare in the \nspms\ population ($\approx0.05$--$0.1\%$). In this population, ECSN-formed NSs have LCs of up to $\approx17\,\msun$, whereas CCSN-formed NSs have LCs of only up to $\approx4.5\,\msun$.

\subsection{Gaia observable \nslc\ Population}
\label{sec:gaia-observed-pop}
\begin{figure*}
    \plotone{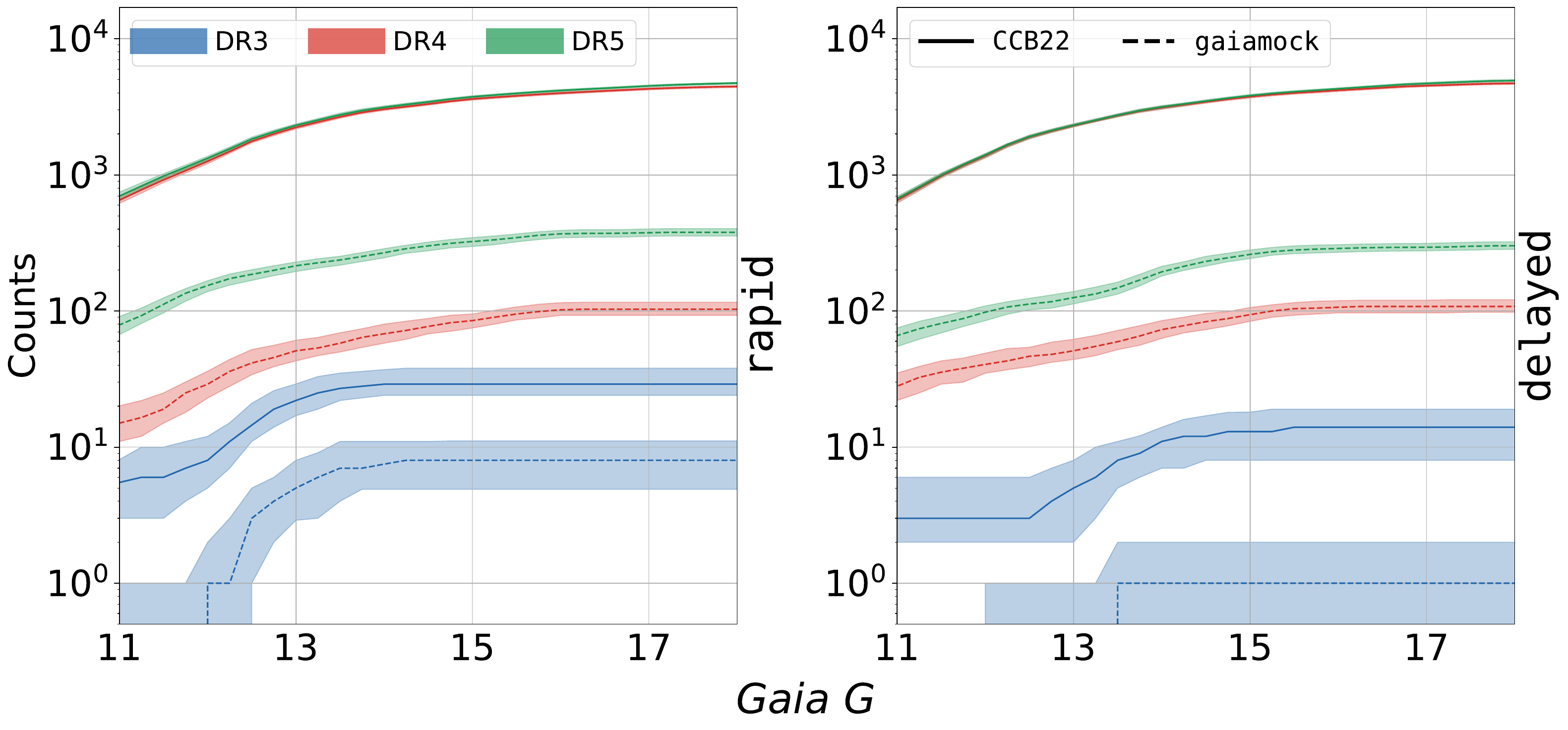}
    \caption{Cumulative number of \gaia-detectable \nslc\ binaries as a function of \gaia\ $G$ magnitude for the \rapid\ (left) and \delayed\ (right) SN models. Colors denote DR3, DR4, and DR5; solid and dashed lines correspond to the \ccb\ and \gaiamock\ selection functions, respectively. Lines show the median and shaded regions the 10th--90th percentile range across the 100 Galactic realizations.}
    \label{fig:detectable-binaries}
\end{figure*}

\autoref{fig:detectable-binaries} shows the cumulative number of \gaia-detectable \nslc\ binaries in DR3, DR4, and DR5 for the \gaia\ selection functions of \citet[][]{Chawla2022} and \gaiamock. With the \ccb\ selection function, we predict $29^{+9}_{-5}\ (14^{+5}_{-6})$, $4536^{+59}_{-90}\ (4738^{+80}_{-80})$, and $4843^{+71}_{-84}\ (5021^{+89}_{-74})$ detectable \nslc\ binaries in DR3, DR4, and DR5, respectively, for the \rapid\ (\delayed) model. In contrast, \gaiamock\ predicts between a factor of a few and more than an order of magnitude fewer detections: $8^{+3}_{-3}\ (1^{+1}_{-1})$, $103^{+13}_{-10}\ (108^{+13}_{-10})$, and $378^{+24}_{-22}\ (301^{+20}_{-17})$ for DR3, DR4, and DR5, respectively, for the \rapid\ (\delayed) model.

The predicted numbers of detectable \nslc\ binaries depend only weakly on the adopted SN prescription. With \gaiamock, the DR4 median of each model lies within the 10th--90th percentile range of the other, while in DR5 the \rapid\ model yields $\approx25\%$ more detections than the \delayed\ model, mainly because it produces about twice as many \nsms\ binaries with $\alpha/\sigma_\xi>20$, nearly all of which \gaiamock\ detects. However, this difference may be comparable to the uncertainties arising from binary evolution assumptions not varied in this work, which are not captured by our realization-to-realization scatter. Hence, the number of \gaia-detectable \nslc\ binaries alone is unlikely to discriminate between the two SN prescriptions.

Both selection functions predict a steep increase in the number of detectable \nslc\ binaries from DR3 to DR4: by more than two orders of magnitude with the \ccb\ selection function, and by one (\rapid) to two (\delayed) orders of magnitude with \gaiamock. This increase primarily reflects the longer observational baseline of DR4 (66 months, compared with 34 months for DR3), which extends the range of orbital periods that can be fully sampled, together with the stringent quality cuts applied to the DR3 astrometric orbital solutions.

The number of detectable \nslc\ binaries predicted with the \ccb\ selection function is comparable to the $27$ \nslc\ candidates identified in \gaia\ DR3 \citep[][]{El-badry_2024a, El-badry_2024b, El-badry_2026}, with the observed number lying between our \rapid\ and \delayed\ predictions. In \citet[][]{Chawla_2025}, we also showed that the orbital properties (e.g., $\ecc$ and $\porb$) of our modeled population agree well with those of the \nslc\ candidates. The much smaller numbers predicted by \gaiamock\ arise because it only attempts an orbital solution if the orbital motion raises the renormalized unit weight error (RUWE) above a threshold. In DR4 and DR5, more than $90\%$ of the \citet[][]{Chawla2022}-detectable binaries fail this first step and are fit with single-star solutions. For the dominant \nsms\ population, these binaries have median orbital periods of $\approx100\,{\rm d}$, distances of $\approx4\,{\rm kpc}$, and astrometric signatures only $\sim2$ times $\sigma_\xi$, compared with $\approx600$--$900\,{\rm d}$, $\approx1.6$--$2.7\,{\rm kpc}$, and $\alpha/\sigma_\xi\approx13$--$20$ for the \gaiamock-detectable binaries. In effect, \gaiamock\ requires $\alpha\gtrsim10\,\sigma_\xi$, for two reasons: the RUWE is computed from CCD-level observations \citep[][]{Lindegren_2021}, whose along-scan uncertainties are $\sim5$ times larger than $\sigma_\xi$, and $\alpha$ is defined in \citet[][]{Chawla2022} as the angular size of the projected orbit, which is about twice the angular semi-major axis of the photocenter orbit. Since this requirement is set by the RUWE threshold, our \gaiamock\ predictions depend strongly on the adopted thresholds \citep[][]{Guerriero_2026}: lower thresholds would allow more binaries to receive orbital solutions and could substantially increase the number of detections. In DR3, the cuts of the \ccb\ selection function already restrict the \nsms\ sample to binaries with $\alpha/\sigma_\xi>10$, but $50$--$66\%$ of these receive an orbital solution that then fails the DR3 quality cuts.

Since \gaiamock\ emulates the DR3 astrometric pipeline, its shortfall relative to the $27$ observed candidates potentially suggests that isolated binary evolution models under-produce detached \nsms\ binaries in au-scale orbits. This may reflect uncertainties in binary evolution that we do not explore here: detailed stellar models find MT to be stable over a wider range of mass ratios than the critical mass ratios adopted in rapid population synthesis codes such as \cosmic\ \citep[][]{Ge_2020, Temmink_2023}, CE evolution remains poorly constrained \citep[][]{Ivanova2013, Thai_2026}, particularly at the extreme mass ratios required to form these systems \citep[][]{Shariat_2026}, and NS natal kicks are uncertain \citep[see][for a review]{Popov_2025}. For example, increasing $\alpha_{\rm CE}$ from 1 to 3 raises the predicted number of detached NS binaries with low-mass MS companions in the MW by a factor of $\sim3$ \citep[][]{Afonina_2026}. These uncertainties are reflected in the substantial differences between population synthesis predictions: using the detailed binary models of BPASS, \citet[][]{Eastep_2026} predict $\sim1.5\times10^{6}$ detached NS binaries with stellar companions in the MW, about an order of magnitude more than \cosmic-based estimates \citep[][]{Chawla_2025, Afonina_2026}, primarily because their weaker CE prescription produces fewer mergers.

In addition, other formation channels, such as formation in triples \citep[][]{Fragione_2020} and in young massive, open \citep[][]{Kamlah_2022, Tanikawa_2024, Banerjee_2026, Iwakura_2026}, and globular clusters \citep[][]{Ye_2019, Fantoccoli_2025}, may also produce \nslc\ binaries and help fill the gap between our predicted detections and the observed candidates. In long-term $N$-body simulations of young massive and open clusters, a majority of the NS--MS binaries with low-mass companions that escape into the Galactic field are dynamically paired, and these encompass the \gaia\ candidates in NS mass, companion mass, and eccentricity, although they also extend to orbital periods beyond $10^{3}$\,d, where \gaia\ may not yet have detected them \citep[][]{Banerjee_2026}. Earlier simulations found that NSs rarely acquire companions dynamically \citep[][]{Tanikawa_2024, Fantoccoli_2025}, which \citet[][]{Iwakura_2026} partly attribute to their focus on the early phase of cluster evolution, when BHs dominate the strong encounters in the cluster core; once the BHs are depleted, \gaia-like NS--MS binaries can form dynamically. Globular clusters, particularly core-collapsed ones, also efficiently form NS binaries \citep[][]{Ye_2019}, although fully disrupted dense clusters are expected to contribute essentially no detectable \nslc\ binaries in DR3 or DR4 \citep[][]{Schiebelbein-Zwack_2026a}.

Taken together, the \ccb\ and \gaiamock\ selection functions provide an optimistic and a more realistic estimate, respectively, of the number of \nslc\ binaries from isolated binary evolution that \gaia\ can detect. For comparison, \citet[][]{Shariat_2026} inferred the intrinsic population of NSs in au-scale orbits with solar-type companions from the 27 DR3 candidates and forward-modeled it through \gaiamock. They forecast $\sim280$--$490$ \nslc\ candidates in DR4 for pessimistic, DR3-like quality cuts and $\sim430$--$650$ for optimistic DR4 cuts (the same as for our \gaiamock\ predictions), with the range in each case set mainly by the poorly constrained orbital-period distribution beyond $\porb\sim1000\,{\rm d}$. This forecast lies between our \gaiamock\ ($\sim100$) and \ccb\ ($\sim4500$--$4700$) estimates for DR4. The DR4 \nss\ catalog will therefore provide a direct test of these predictions: a number of \nslc\ candidates well above $\sim100$ would indicate either a substantial contribution from formation channels beyond isolated binary evolution, or that the effective \gaia\ selection function is less restrictive than that modeled by \gaiamock, for example through a lower RUWE threshold for attempting orbital solutions.

\subsection{Key Properties of the \gaia-Observable \nslc\ Population}
\label{sec:detectable-prop}

\begin{figure*}
    \plotone{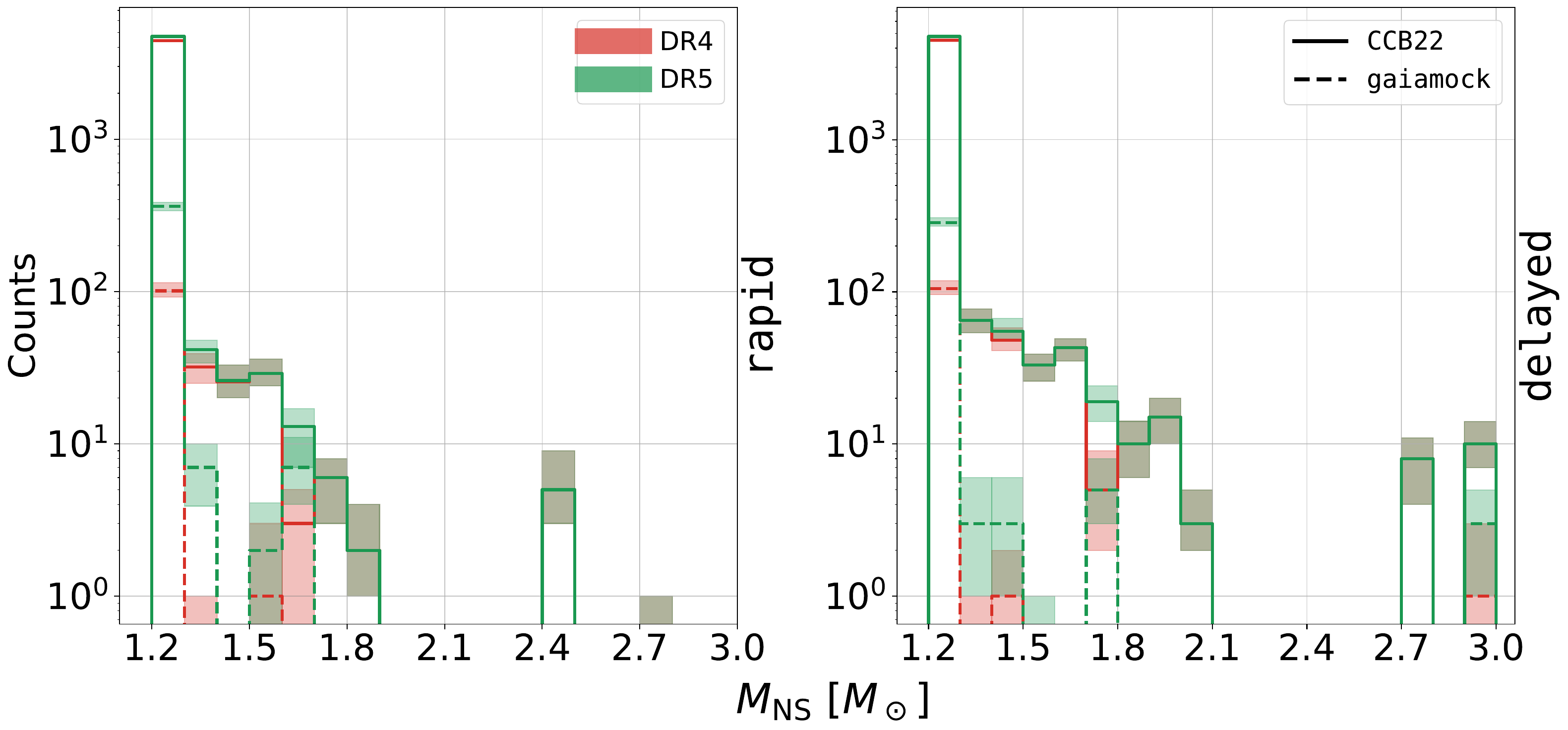}
    \caption{NS mass distribution of \gaia-detectable \nslc\ binaries (MS and PMS companions combined) for the \rapid\ (left) and \delayed\ (right) SN models. Colours denote DR4 and DR5; solid and dashed lines correspond to the \ccb\ and \gaiamock\ selection functions, respectively. Lines show the median and shaded regions the 10th--90th percentile range across the 100 Galactic realisations.}
    \label{fig:detectable-mns}
\end{figure*}

In this section, we highlight the key properties of the \gaia-detectable \nslc\ population in the upcoming DR4 and DR5. We have already discussed the DR3-detectable population in \citet[][]{Chawla_2025}, where we showed that both the number and the orbital properties of the DR3 \nslc\ candidates agree with our modelled population. In addition, the number of DR3 detections predicted by \gaiamock\ is too small to yield statistically meaningful conclusions.

In \autoref{sec:intrinsic-prop}, we showed that, although the $\mns$ distributions of the entire \nslc\ population are similar for the two SN models, those of the detached short-period ($\porb/\yr\le10$) \nsms\ binaries differ clearly: the \rapid\ model produces NSs only up to $\sim2.1\,\msun$, whereas in the \delayed\ model $\mns$ extends up to $3\,\msun$ (\autoref{fig:intrinsic-mns}). This difference, however, is largely absent in the \gaia-detectable population, for which the $\mns$ distributions are similar for the \rapid\ and \delayed\ models (\autoref{fig:detectable-mns}). With the \ccb\ selection function, the DR4 and DR5 distributions are nearly identical, since the numbers of detectable binaries in the two data releases are similar. With \gaiamock, the $\mns$ distribution peaks between $1.2$ and $1.3\,\msun$ in both DR4 and DR5, and only a small fraction ($\sim1\%$) of \nslc\ binaries have $\mns>1.5\,\msun$. This peak coincides with the narrow mass range of NSs formed via ECSN, which, owing to their weaker natal kicks, survive in binaries more often than CCSN-formed NSs and account for roughly two-thirds of the intrinsic detached \nsms\ population in both SN models (\autoref{sec:intrinsic-prop}), suggesting that the detectable population is dominated by ECSN-formed NSs. Since the masses of NSs formed via ECSN do not depend on the adopted CCSN engine, the differences between the \rapid\ and \delayed\ models are largely washed out in the detectable population. Notably, this occurs even though the astrometric signal of an \nslc\ binary increases with the NS mass at fixed orbital period and LC mass, which favours the detection of binaries hosting more massive NSs. That the detectable $\mns$ distributions nevertheless remain dominated by NSs with $1.2$--$1.3\,\msun$ indicates that this selection effect is weaker than the intrinsic dominance of low-mass, ECSN-formed NSs. We note that, as discussed in \citet[][]{Chawla_2025}, our modelled NS masses are systematically lower than those inferred for the \gaia\ DR3 NS candidates, which may be due to the assumed NS birth-mass prescription. A similar offset has been reported in studies of double neutron stars and NS X-ray binaries \citep[][]{Andrews_2015}.

\nslc\ binaries hosting massive NSs ($\mns\ge2.1\,\msun$) originate from \nspms\ binaries in which the NS has accreted mass from its companion via RLOF, consistent with the detached short-period \nspms\ binaries covering the full $\mns$ range up to $3\,\msun$ in both SN models (\autoref{sec:intrinsic-prop}). A small fraction of these \nspms\ binaries are detectable with the \ccb\ selection function in both SN models, whereas with \gaiamock\ they appear only in the \delayed\ population. In this case, all of them have $\mns>2.5\,\msun$ and account for $\sim1\%$ of the \gaiamock-detectable population in both DR4 and DR5. Such systems are of particular interest, as they are heavier than the most massive NSs measured to date \citep[][]{Romani_2022} and thus probe the maximum NS mass and dense-matter EoS. Moreover, NSs with $2.5\lesssim\mns/\msun\le3$ may be observationally indistinguishable from low-mass BHs.

Therefore, the features that distinguish the intrinsic $\mns$ distributions of the two SN models, most notably the extended high-mass tail in the \delayed\ model, cannot be distinctly recovered in the \gaia-observable population. This contrasts with the \bhlc\ population, for which the lower mass gap produced by the \rapid\ model remains imprinted on the \gaia-detectable population \citep[][]{Chawla2022}. Identifying these features in the \nslc\ population will therefore require larger samples and more precise mass measurements, which may become possible with future astrometric missions and dedicated spectroscopic follow-up. Since most \nslc\ binaries in our models form through CE evolution, such measurements would constrain not only NS formation physics but also binary evolution, in particular the CE phase.

\begin{figure*}
    \plotone{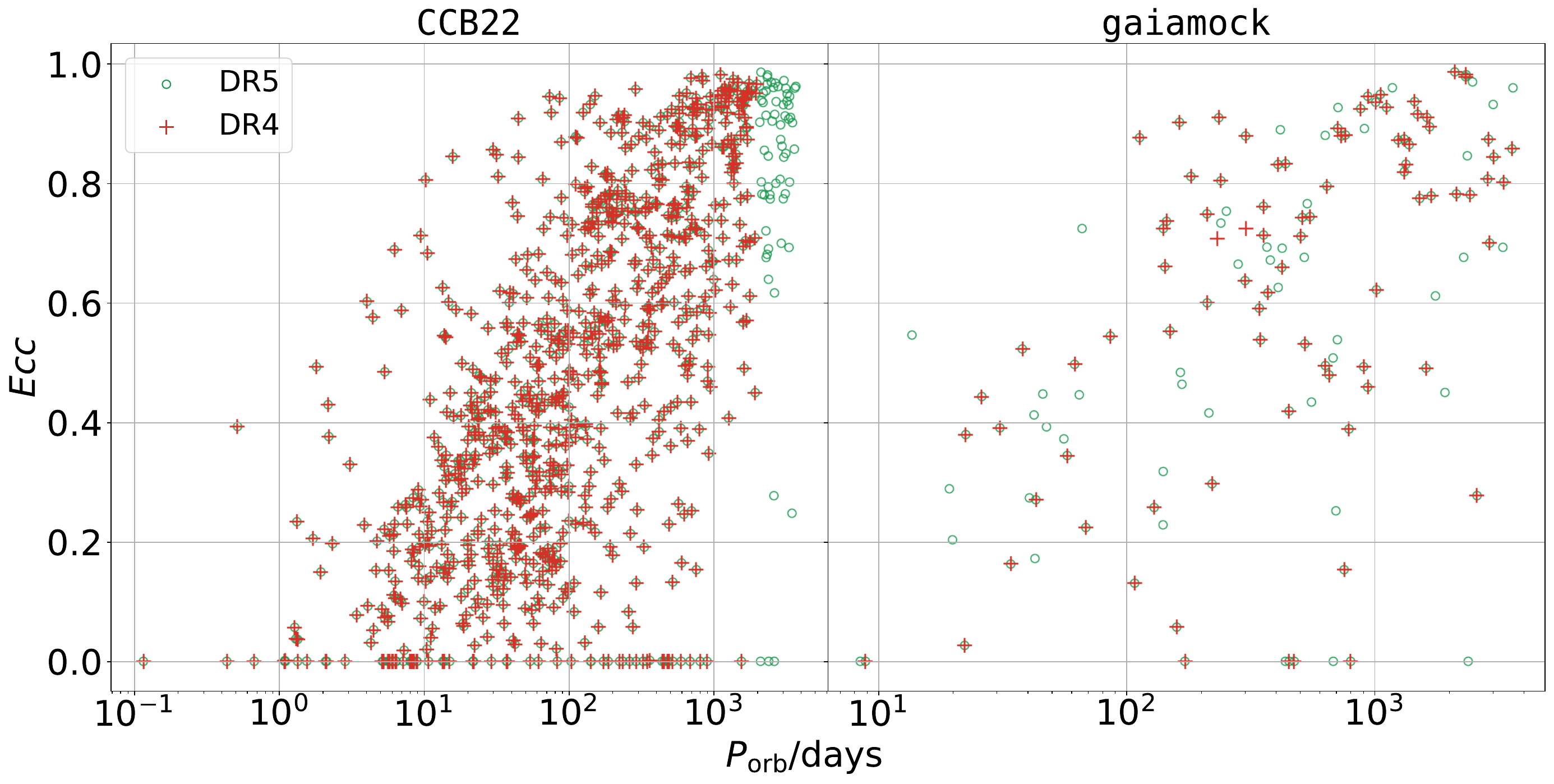}
    \caption{Orbital period--eccentricity distribution of \gaia-detectable \nslc\ binaries (MS and PMS companions combined) for the \rapid\ SN model, combining all 100 Galactic realizations. Left and right panels show binaries detectable with the \ccb\ and \gaiamock\ selection functions, respectively. Green circles and red plus signs denote DR5 and DR4 detections, respectively. The corresponding figure for the \delayed\ SN model is shown in \autoref{fig:porb-ecc-delayed} in the Appendix.}
    \label{fig:porb-ecc-rapid}
\end{figure*}

The orbital properties ($\ecc$ and $\porb$) of the \gaia-detectable \nslc\ population are largely similar in DR4 and DR5 (\autoref{fig:porb-ecc-rapid}; the \delayed\ model shows similar distributions). The two selection functions, however, sample different parts of the $\porb$ distribution: the median $\porb$ of the detectable \nsms\ binaries is $\approx100\,{\rm d}$ with the \ccb\ selection function but $\approx600$--$900\,{\rm d}$ with \gaiamock, since most short-period binaries have astrometric signatures too small to raise the RUWE above the threshold for an orbital solution (\autoref{sec:gaia-observed-pop}). As discussed in \autoref{sec:intrinsic-prop}, the present-day $\ecc$ of \nslc\ binaries is set by the SN kicks and by subsequent circularization through tides and post-SN MT/CE episodes. \nslc\ binaries in tight orbits are circularized by tides, an effect that weakens for wider orbits, which results in a positive correlation between $\ecc$ and $\porb$. The SN kicks further strengthen this correlation: for a given kick, the induced eccentricity increases with decreasing orbital velocity, i.e., with increasing orbital separation. As in the intrinsic population, the fraction of circular ($\ecc=0$) binaries is larger among \nspms\ binaries, most of which experienced a second MT/CE episode after NS formation. With the \ccb\ selection function, fewer than $1\%$ of the detectable \nsms\ binaries have circular orbits, whereas $48$--$50\%$ of the detectable \nspms\ binaries are circular in DR5, rising to $53$--$54\%$ in DR4, depending on the SN model. These fractions are lower than in the intrinsic population, where $3.5\%$ of the \nsms\ and $91$--$92\%$ of the \nspms\ binaries are circular (\autoref{sec:intrinsic-prop}). Circular \nspms\ binaries that underwent stable MT after NS formation may host NSs that accreted mass and were spun up, making them promising targets for radio pulsar searches.

With \gaiamock, the circular fraction of the detectable \nspms\ binaries depends strongly on the SN model, ranging from $43\%$ (DR4) and $31\%$ (DR5) for the \rapid\ model to $82\%$ (DR4) and $70\%$ (DR5) for the \delayed\ model. This strong model dependence arises because the \gaiamock-detectable \nspms\ binaries originate from only a handful of distinct binaries in our synthetic populations ($15$--$19$ in DR5), each of which recurs across the Galactic realizations. Consequently, the \gaiamock-detectable \nspms\ binaries of the two models also differ in other properties; in DR5, for example, their median $\porb$ is $\approx750$ and $\approx350\,{\rm d}$ and their median LC mass is $\approx2.2$ and $\approx3.5\,\msun$ for the \rapid\ and \delayed\ models, respectively. Nevertheless, two trends hold in both SN models. All \ccb-detectable \nspms\ binaries with $\porb\lesssim5\,{\rm d}$ are circular, and none of them is detected with \gaiamock, even in DR5. At $\porb\gtrsim250\,{\rm d}$, where most \gaiamock\ detections lie, the ratio of \gaiamock\ to \ccb\ detections is $0.23$ for circular binaries and $0.26$ for eccentric ($\ecc>0$) binaries in the \rapid\ model, and $0.25$ and $0.07$, respectively, in the \delayed\ model. Circular binaries are thus retained about as often as eccentric ones in the \rapid\ model and more often in the \delayed\ model, consistent with the decreasing sensitivity of \gaia\ to highly eccentric orbits \citep[][]{El-badry_2024d}. Hence, \gaiamock\ does not preferentially select eccentric binaries; the lower circular fraction of its \rapid\ sample instead reflects the loss of short-period binaries, which are almost all circular.
% \sourav{[Do you mean gaiamock preferrentially under select eccentric binaries, especially at longer Porb/Gaia observation duration?]}

With the \ccb\ selection function, the inclusion of binaries with $5.5<\porb/\yr<10$ in DR5 adds wide, mostly highly eccentric binaries to the detectable population. In contrast, with \gaiamock\ the DR5 population differs from the DR4 one mainly in size, with only a moderate shift toward longer orbital periods (median $\porb\approx790$--$880\,{\rm d}$, compared with $\approx610$--$650\,{\rm d}$ in DR4, for the \nsms\ binaries). The $\ecc$--$\porb$ distribution of the $\sim100$ (DR4) to $\sim300$--$380$ (DR5) \nslc\ binaries expected from \gaiamock\ will therefore help constrain the relative contributions of low-kick (ECSN and AIC) and high-kick (CCSN) NS formation, as well as the role of post-SN MT/CE episodes.

\section{Comparison with Known Pulsars in the ATNF Pulsar catalog}
\label{sec:atnf}

\begin{figure}
    \plotone{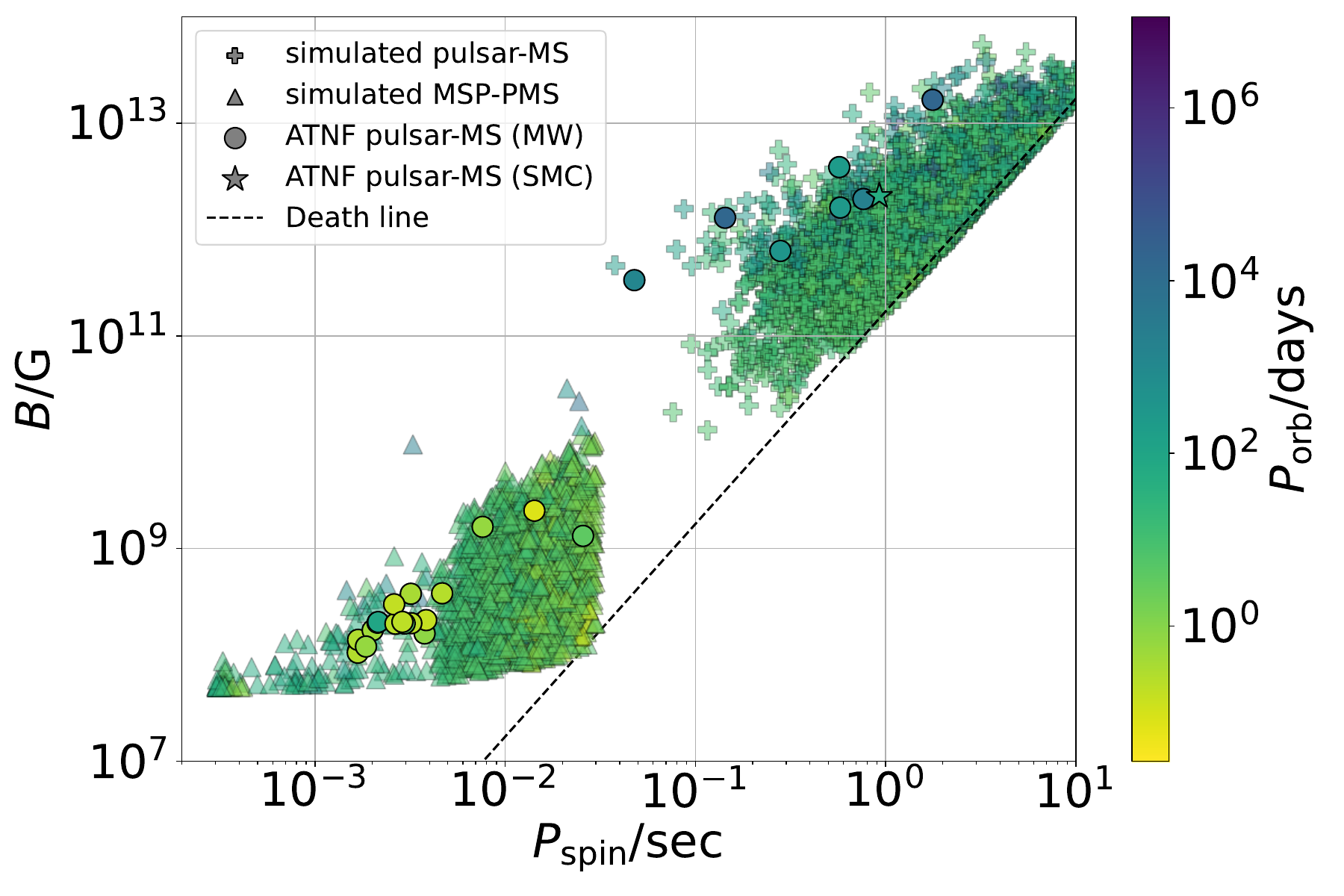}
    \caption{Surface magnetic field strength versus spin period. Filled plus signs and triangles show the simulated pulsars in detached \nsms\ binaries and the simulated MSPs in detached \nspms\ binaries, respectively, in the intrinsic population for the \rapid\ SN model. Circles and the star show the ATNF pulsars with MS companions, including MSPs, in the MW field and in the SMC (PSR~J0045$-$7319), respectively; pulsars in globular clusters are excluded. The color of each symbol indicates the $\porb$ of the binary, shown on a logarithmic scale in the color bar. The dashed line shows the pulsar death line, $B/P_{\rm spin}^2=0.17\times10^{12}\,{\rm G\,s^{-2}}$. The corresponding figure for the \delayed\ SN model is shown in \autoref{fig:atnf-B-P-delayed} in the Appendix.}
    \label{fig:atnf-B-P}
\end{figure}

The Australia Telescope National Facility (ATNF) Pulsar catalog \citep[v2.9.0;][]{Manchester_2005} lists 4503 known pulsars, of which 557 are in binaries and 52 have MS companions. For each pulsar, the catalog lists the measured timing parameters, including the spin period ($P_{\rm spin}$), its time derivative ($\dot{P}_{\rm spin}$), and the parallax where available, and, for binary pulsars, orbital parameters such as $\porb$ and $\ecc$. It also lists quantities derived from these measurements, including the surface magnetic field strength ($B$), estimated from $P_{\rm spin}$ and $\dot{P}_{\rm spin}$ assuming magnetic dipole braking, and estimates of $\mlc$ from the binary mass function. Since $\mlc$ cannot be measured from pulsar timing alone, we adopt the median companion mass, which assumes an orbital inclination of $60^\circ$ and an NS mass of $1.35\,\msun$. The 52 pulsars with MS companions fall into three groups. Fifteen reside in globular clusters, where they likely formed through dynamical interactions and where their measured $\dot{P}_{\rm spin}$ can be biased by acceleration in the cluster potential. A further 29 are field MSPs, most of which are redbacks with $\porb<1.1\,{\rm d}$. The remaining eight are young pulsars with massive B- or Be-star companions in wide, eccentric orbits, one of which (PSR~J0045$-$7319) resides in the Small Magellanic Cloud (SMC). We exclude the globular-cluster pulsars from our comparison, since our models do not include dynamical formation channels, but retain PSR~J0045$-$7319, which likely formed through isolated binary evolution, albeit at the lower metallicity of the SMC. Of the remaining 37 pulsars, 29 have a measured positive $\dot{P}_{\rm spin}$, and hence an estimate of $B$, and are shown in \autoref{fig:atnf-B-P}.

In our simulated population, each NS is assigned at birth a magnetic field strength and a spin period drawn randomly from the ranges $10^{11.5}$--$10^{13.8}\,{\rm G}$ and $30$--$1000\,{\rm ms}$, respectively, consistent with the observed population of young pulsars \citep[][]{Manchester_2005}. \cosmic\ then evolves the NS spin and magnetic field using the prescriptions of \citet[][]{Hurley2002} and \citet[][]{Kiel_2008}. For isolated NSs and NSs in detached binaries, magnetic dipole radiation dominates the spin evolution, and treating NSs as solid spheres gives a spin-down rate of
\begin{equation}
    \dot{P}_{\rm spin} = K\,\frac{B^2}{P_{\rm spin}},
    \label{eq:spindown}
\end{equation}
where $K=9.87\times10^{-48}\,{\rm yr\,G^{-2}}$. The magnetic field decays exponentially with the age of the NS, $T$, as
\begin{equation}
    B = B_0\exp\left(-\frac{T}{\tau}\right),
    \label{eq:bdecay}
\end{equation}
where $B_0$ is the birth field strength and $\tau=3\,{\rm Gyr}$ \citep[][]{Kiel_2008}. This long timescale implies only weak field decay over the lifetimes of radio pulsars, in line with the finding of \citet[][]{Faucher-Giguere_2006} that isolated radio pulsars show no significant field decay over $\sim100\,{\rm Myr}$, and with magneto-thermal models of NS field evolution, in which the fields of normal pulsars decay only modestly, by a factor of $\sim2$, while substantial decay is confined to magnetar-strength fields \citep[e.g.,][]{Popov_2010, Vigano_2013}.

In non-detached binaries, in contrast, accretion can substantially alter the spin period and magnetic field of the NS on short timescales. During accretion through RLOF, we assume that the accreted material buries the magnetic field \citep[e.g.,][]{Bhattacharya_1991, Rappaport_1995, Kiel_2008, Tauris_2012}, which then decays as
\begin{equation}
    B = \frac{B_0}{1+\Delta M/(10^{-6}\,\msun)}
        \exp\left(-\frac{T-t_{\rm acc}}{\tau}\right),
    \label{eq:bdecay-acc}
\end{equation}
where $\Delta M$ is the accreted mass and $t_{\rm acc}$ is the duration of the accretion phase. At the same time, the NS is spun up by the angular momentum carried by the accreted material \citep[][their Equation~54]{Hurley2002}. Since NSs are assumed not to accrete during CE phases \citep[][]{Hurley2002}, only stable MT can recycle NSs into MSPs in our models. Other processes, such as wind mass loss, tidal evolution, magnetic braking, and SN kicks, have only minor effects on the evolution of $B$ and $P_{\rm spin}$.

We adopt the standard pulsar death line, $B/P_{\rm spin}^2=0.17\times10^{12}\,{\rm G\,s^{-2}}$ \citep[][]{Ruderman_1975, Bhattacharya_1992}, although its exact location remains under debate \citep[e.g.,][]{Zhang_2000, Zhang_2003, Zhou_2017}. We define a pulsar as any NS lying above the death line, an MSP as any pulsar with $P_{\rm spin}<30\,{\rm ms}$, and a young pulsar as any pulsar that is not an MSP. Since accretion onto the NS is expected to quench its radio emission, we only consider pulsars in detached binaries and, to match the ATNF comparison sample, focus on \nsms\ binaries. For the \rapid\ (\delayed) model, our intrinsic population contains $26{,}685^{+217}_{-194}\ (27{,}382^{+262}_{-160})$ pulsars in detached \nsms\ binaries, all of which are young pulsars. In our models, all NSs with MS companions that satisfy the MSP criterion are currently accreting from their companions via RLOF, as also found by \citet[][]{Kiel_2008}, whereas the detached MSPs all have PMS companions. The accreting systems would appear as low-mass X-ray binaries (LMXBs) or transitional MSPs rather than as radio MSPs, and are therefore not shown in \autoref{fig:atnf-B-P} and \autoref{fig:atnf-B-P-delayed}, which instead include the detached MSPs with PMS companions for comparison. Our models thus produce no detached MSPs with MS companions, whereas the 29 field MSPs with MS companions in the ATNF catalog are radio pulsars, almost all of which reside in tight orbits ($\porb\lesssim7\,{\rm d}$), most of them as redbacks. PSR~J1903+0327, which resides in a wide, eccentric orbit ($\porb\approx95\,{\rm d}$, $\ecc=0.44$), likely formed in a former triple system \citep[][]{Freire_2011}, a channel not included in our models. The origin of these radio MSPs with MS companions in tight orbits therefore remains an open question, which we leave to future work.

\autoref{fig:atnf-B-P} compares the $B$--$P_{\rm spin}$ distribution of the simulated pulsars in detached \nsms\ binaries with that of the ATNF pulsars with MS companions. The simulated young pulsars agree well with the young ATNF pulsars, not only in $B$ and $P_{\rm spin}$ but also in $\porb$. In the \rapid\ model, they also predominantly have massive companions, with a median $\mlc$ of $5.5\,\msun$ (10th--90th percentile range of $3.2$--$10.1\,\msun$), consistent with the B- and Be-star companions of the young ATNF pulsars. Indeed, all MS companions identified around young pulsars are massive OB stars \citep[][]{Kaspi_1996, Bassa_2011, Shannon_2024}. In our models, such binaries survive either because their massive LCs emerge from CE in wide orbits that withstand the weak ECSN kicks, or because they avoid CE through stable MT, during which the LC accretes mass from the NS progenitor and is rejuvenated (\autoref{sec:intrinsic-prop}). Possible exceptions are PSR~B1820$-$11 and PSR~J1954+2529, young pulsars with roughly solar-mass companions in eccentric orbits \citep[][]{Phinney_1991, Parent_2022}, although these companions are undetected and may instead be massive WDs \citep[][]{El-badry_2024b}.

Unlike the young pulsars, the ATNF MSPs, which occupy $P_{\rm spin}<30\,{\rm ms}$ and $B\approx10^{8}$--$2\times10^{9}\,{\rm G}$, have no counterparts among our simulated pulsars in detached \nsms\ binaries, since all NSs with MSP-like spin periods and magnetic fields that have MS companions are still accreting from them in our models. Our models do, however, produce detached MSPs whose $B$ and $P_{\rm spin}$ overlap with those of the ATNF MSPs (triangles in \autoref{fig:atnf-B-P}), but all of them have PMS companions. Since PMS stars that are not giants can be difficult to distinguish from MS stars, some of the ATNF MSPs listed with MS companions could be such systems. Indeed, two of them are known to have evolved companions: that of PSR~J1306$-$4035 is slightly evolved off the MS \citep[][]{Swihart_2019}, and that of PSR~J1417$-$4402 is a giant \citep[][]{Strader_2015, Camilo_2016}. Beyond $B$ and $P_{\rm spin}$, recycling also leaves its imprint on the masses and orbits of MSPs. Observed recycled NSs are on average more massive than NSs in BNS systems, with mean masses of $\approx1.54$ and $\approx1.33\,\msun$, respectively \citep[][]{Ozel_2016}, which suggests significant mass accretion during spin-up \citep[][]{Ozel_2012, Kiziltan_2013}. The MSP mass distribution may also be bimodal, with peaks at $\approx1.39$ and $\approx1.81\,\msun$, but this bimodality more likely reflects differences in NS birth masses than in recycling \citep[][]{Antoniadis_2016}. Consistent with this, the \gaia\ DR3 NSs \citep[][]{El-badry_2024a, El-badry_2024b} and the firstborn NSs in Galactic BNS systems, neither of which is expected to have accreted much mass, have strikingly similar mass distributions that extend to high masses, supporting the hypothesis that NSs can be born massive \citep[][]{Schiebelbein-Zwack_2026b}. Moreover, fully recycled MSPs in the Galactic field have nearly circular orbits, with only a few exceptions \citep[][]{Ozel_2016}.

For the \rapid\ (\delayed) model, $22^{+7}_{-5}\ (13^{+5}_{-4})$ and $124^{+14}_{-15}\ (53^{+10}_{-6})$ of the \gaiamock-detectable \nsms\ binaries in DR4 and DR5, respectively, host young pulsars. The two models contain similar numbers of young pulsars in detached \nsms\ binaries, but these \gaiamock-detectable pulsar hosts originate from only a small number of distinct binaries in our synthetic populations ($26$--$28$ in DR5), each of which recurs across the Galactic realizations. Those in the \rapid\ model have wider orbits, more massive companions, and smaller distances (median $\porb\approx1300$ versus $1000\,{\rm d}$, $\mlc\approx5.1$ versus $3.9\,\msun$, and distances of $3.3$ versus $3.9\,{\rm kpc}$), which give them larger astrometric signatures and hence more detections. Since the radio beams of only $\sim10\%$ of normal pulsars sweep across our line of sight \citep[][]{Tauris_1998, Faucher-Giguere_2006}, we expect only $\sim2\ (1)$ and $\sim12\ (5)$ of these pulsars to be beamed toward us in DR4 and DR5, respectively; nevertheless, such systems are prime targets for radio follow-up of the \gaia\ \nslc\ candidates.

Beyond hosting young pulsars today, many \nslc\ binaries are also progenitors of recycled pulsars \citep[e.g.,][]{Bhattacharya_1991, Tauris_2006, Tauris_2023}. Those with Sun-like LCs ($\mlc\sim1\,\msun$) are expected to undergo stable MT onto the NS once the LC evolves off the MS, ending as NS--WD binaries \citep[][]{Tauris_2023, Chattopadhyay_2025, Chattopadhyay_2026}. Depending on whether the orbit circularizes before MT begins, the NS becomes either a fully recycled MSP in a circular orbit or a mildly recycled pulsar in a wide, eccentric orbit \citep[][]{Chattopadhyay_2026}. However, \citet[][]{Chattopadhyay_2026} argue that most known MSP--WD binaries descend from NSs with more massive companions that are recycled during a CE phase. This channel is absent from our models, in which NSs do not accrete during CE. Since intermediate-mass LCs are abundant in our \nsms\ population, including the \gaiamock-detectable pulsar hosts ($\mlc\approx4$--$5\,\msun$), \gaia\ may also reveal the progenitors of these canonical MSP--WD binaries.

\section{Conclusions}
\label{sec:conclusions}

In this paper, we have presented predictions for the population of \nslc\ binaries in the MW that \gaia\ can detect astrometrically. We generated synthetic Galactic populations of \nslc\ binaries with \cosmic, using the star formation history, metallicities, and spatial distribution of stars in the FIRE-2 MW-mass galaxy m12i, and accounted for dust extinction along each line of sight across 100 Galactic realizations. We considered two SN engine prescriptions, the \rapid\ and \delayed\ models of \citet[][]{Fryer2012}, and three NS formation channels (CCSN, ECSN, and AIC), and we assessed the detectability of the resulting binaries in \gaia\ DR3, DR4, and DR5 using both the approximate selection function of \ccb\ and the more realistic \gaiamock\ \citep[][]{El-badry_2024d}. Finally, we compared our models with the known pulsars with MS companions in the ATNF Pulsar catalog. Our main findings can be summarized as follows:
\begin{enumerate}
      \item {\em Most present-day \nslc\ binaries have survived a CE phase.}\\
      About $94\%$ of all present-day \nslc\ binaries, including those currently transferring mass, have experienced a CE (initiated by the NS progenitor), a much larger fraction than for \bhlc\ binaries \citep[$\sim4$--$44\%$;][]{Chawla2022}, because the orbital shrinkage during the CE makes them more likely to survive the strong natal kicks of NSs. Among those with $\porb/\yr\le10$, $\sim93\%$ have experienced a CE and $\sim7\%$ only stable MT. The stronger kicks and lower masses of NSs also disrupt a large fraction of wide \nslc\ binaries, so that, unlike that of \bhlc\ binaries, the $\porb$ distribution of \nslc\ binaries lacks a long-period, non-CE peak and is not bimodal (see \autoref{sec:intrinsic-prop} and \autoref{fig:intrinsic-porb}).
    \item {\em \gaia\ DR4 and DR5 could detect $\sim100$ and $\sim300$ \nslc\ binaries formed through isolated binary evolution.}\\
    With \gaiamock, we predict $\sim100$--$110$ and $\sim300$--$380$ detectable \nslc\ binaries in DR4 and DR5, respectively, across the two SN models. The approximate selection function of \ccb\ predicts $\sim4500$--$4700$ and $\sim4800$--$5000$, providing an optimistic upper bound. This difference arises because more than $90\%$ of the \citet[][]{Chawla2022}-detectable binaries have astrometric signatures too small to raise the RUWE above the threshold beyond which \gaiamock\ attempts an orbital solution. Both selection functions predict a steep increase from DR3 to DR4, driven by the longer observational baseline of DR4 and the stringent quality cuts applied to the DR3 orbital solutions (see \autoref{sec:gaia-observed-pop} and \autoref{fig:detectable-binaries}).
    \item {\em The DR4 \nss\ catalog will test the contribution of formation channels beyond isolated binary evolution.}\\
    While the \citet[][]{Chawla_2025} predictions for DR3 ($\sim14$--$29$) bracket the 27 \nslc\ candidates identified in the DR3 \nss\ catalog, \gaiamock\ predicts only $\sim1$--$8$. This shortfall suggests contributions from other channels, such as formation in star clusters and in triples. A DR4 sample of \nslc\ candidates well above $\sim100$ would indicate a substantial contribution from these channels, or a less restrictive effective selection function than that modeled by \gaiamock\ (see \autoref{sec:gaia-observed-pop}).
    \item {\em The \gaia-detectable \nslc\ population is largely insensitive to the SN prescription.}\\
    With \gaiamock, the \rapid\ and \delayed\ models predict consistent numbers of detections in DR4 and differ by only $\approx25\%$ in DR5, a difference that may be comparable to the uncertainties in binary evolution (see \autoref{sec:gaia-observed-pop}). The two models do differ in their intrinsic $\mns$ distributions: for detached short-period \nsms\ binaries, $\mns$ extends only to $\sim2.1\,\msun$ in the \rapid\ model but up to $3\,\msun$ in the \delayed\ model. However, this difference is not preserved in the \gaia-detectable populations, whose $\mns$ distributions are similar for the two models; with \gaiamock, both peak at $1.2$--$1.3\,\msun$, the mass range of NSs formed via ECSN, whose masses do not depend on the CCSN engine. Hence, neither the number of detections nor their $\mns$ distribution is likely to discriminate between the two SN engines, in contrast to \bhlc\ binaries, for which the lower mass gap of the \rapid\ model remains imprinted on the \gaia-detectable population (see \autoref{sec:detectable-prop} and \autoref{fig:detectable-mns}).
    \item {\em \gaia\ may detect NSs heavier than the most massive known pulsars.}\\
    In the \delayed\ model, $\sim1\%$ of the \gaiamock-detectable binaries host NSs with $\mns>2.5\,\msun$, which grew through accretion from their PMS companions. Such systems would probe the maximum NS mass and the dense-matter EoS, although at these masses they may be difficult to distinguish from low-mass BHs (see \autoref{sec:detectable-prop}).
    \item {\em Circular \gaia-detectable \nspms\ binaries are promising targets for radio pulsar searches.}\\
    The present-day $\ecc$ of \nslc\ binaries is set by SN kicks and subsequent circularization, resulting in a positive correlation between $\ecc$ and $\porb$. Roughly half of the \citet[][]{Chawla2022}-detectable \nspms\ binaries are circular; most of these experienced MT/CE after NS formation, so their NSs may have been spun up. \gaiamock\ does not preferentially select eccentric binaries, but it detects none of the circular \nspms\ binaries with $\porb\lesssim5\,{\rm d}$ (see \autoref{sec:detectable-prop} and \autoref{fig:porb-ecc-rapid}).
   \item {\em Our model pulsars match the young ATNF pulsars with MS companions, but not the MSPs.}\\
   The pulsars in our detached \nsms\ binaries are all young and agree well with the young ATNF pulsars with MS companions in $B$, $P_{\rm spin}$, and $\porb$. None of them is an MSP, since all NSs with MSP-like spin periods and MS companions are still accreting in our models; the observed MSPs with MS companions are instead shaped by processes not included in \cosmic, such as irradiation and ablation of the companion by the pulsar wind and formation in triple systems. Across the two SN models, $\sim13$--$22$ and $\sim50$--$120$ of the \gaiamock-detectable \nsms\ binaries in DR4 and DR5, respectively, host young pulsars, making them targets for radio follow-up (see \autoref{sec:atnf} and \autoref{fig:atnf-B-P}).
\end{enumerate}
Overall, \gaia-detectable \nslc\ binaries provide a unique window into NS formation, SN kicks, and CE evolution. \gaia\ has already begun to reveal this population, with $27$ \nslc\ candidates identified in DR3 \citep[][]{El-badry_2024a, El-badry_2024b, El-badry_2026}. The size of the DR4 sample will provide a direct test of our predictions: with \gaiamock, isolated binary evolution yields only $\sim100$ detections, whereas the population inferred from the DR3 candidates is forecast to yield $\sim300$--$650$ \citep[][]{Shariat_2026}. Beyond \gaia, time-domain surveys such as the Zwicky Transient Facility \citep[ZTF;][]{Bellm_2018} and the Vera C. Rubin Observatory Legacy Survey of Space and Time \citep[LSST;][]{Ivezic_2019} could uncover shorter-period and fainter \nslc\ binaries through their photometric variability \citep[][]{Wiktorowicz2021}. Combined with \gaia, the deep, wide-field epoch astrometry of the Chinese Space Station Survey Telescope \citep[CSST;][]{CSST_2026} could extend astrometric searches to fainter magnitudes and longer orbital periods \citep[][]{Wen_2026}, while the near-infrared Galactic Bulge Time Domain Survey of the recently launched Nancy Grace Roman Space Telescope \citep[][]{ROMAN_2015} could detect from $\mathcal{O}(10)$ to a few hundred detached \nslc\ binaries in the crowded, dusty bulge, where \gaia's astrometry degrades \citep[][]{Lam_2026b, Eastep_2026}. Ground-based multi-epoch spectroscopic surveys will be crucial for confirming the \gaia\ \nslc\ candidates and measuring their masses \citep[][]{Li_2026b, Lam_2026a, Simon_2026, Zhou_2026}, while cross-matching radio pulsar surveys can reveal a pulsar detection confirming the NS nature of the dark companion. Moreover, since CE evolution and NS natal kicks also govern the formation of the BNS and NS--BH mergers detected by the LVK \citep[e.g.,][]{vanSon_2025, Chattaraj_2026} and of the sGRBs associated with them \citep[][]{Fishbach_2026, Kunnumkai_2026}, the \gaia\ \nslc\ population provides an independent test of the binary physics underlying these transients. With the upcoming \gaia\ DR4 and DR5, together with these complementary surveys, we will soon have a rich sample of \nslc\ binaries with which to test our predictions and constrain the underlying binary physics.

\section*{Acknowledgement}

CC acknowledges support from ARCO. SC acknowledges support from the Department of Atomic Energy, Government of India, under project no. 12-R\&D-TFR-5.02-0200 and RTI 4002. KB acknowledges support from the Falco DeBenedetti Career Development Professorship. All simulations were done using cloud computing on Azure.

\software{astropy \citep{astropy:2013, astropy:2018, Astropy_2022}
          \cosmic\ \citep{Breivik_2020, Breivik_2021}; \mwdust\ \citep{Bovy2016}; \texttt{isochrones}\ \citep{Isochrones_2015}; \texttt{matplotlib}\ \citep{matplotlib}; \texttt{numpy}\ \citep{numpy}; \texttt{scipy}\ \citep{SciPy-NMeth_2020}; \texttt{pandas}\ \citep[][]{Pandas_2010}}; \gaiamock\ \citep[][]{El-badry_2024d}

% \clearpage
\bibliography{references}{}
\bibliographystyle{aasjournal}
\appendix
\restartappendixnumbering
\section{Supplementary figures}

Here we present selected results from our \delayed\ model.
\begin{figure*}[h!]
    \plotone{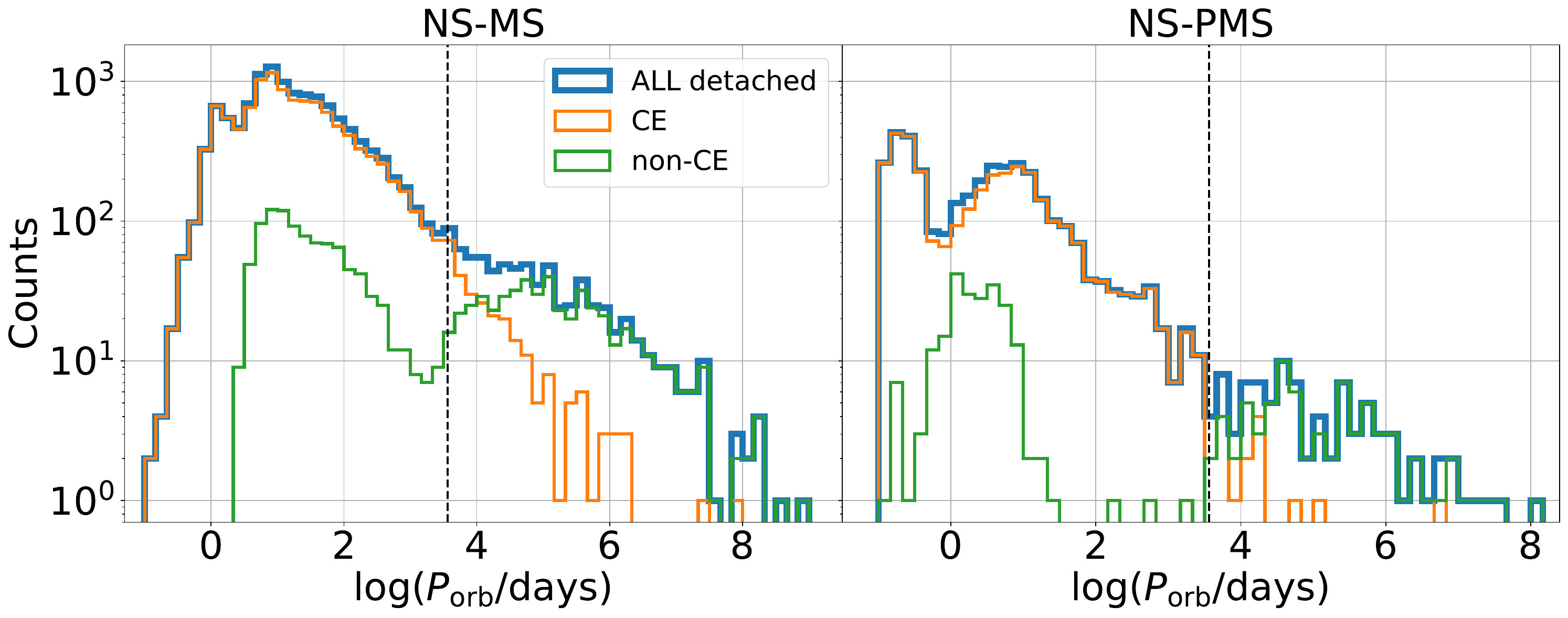}
    \caption{Same as \autoref{fig:intrinsic-porb} but for the \delayed\ SN model.}
    \label{fig:intrinsic-porb-delayed}
\end{figure*}

\begin{figure}[h!]
    \plotone{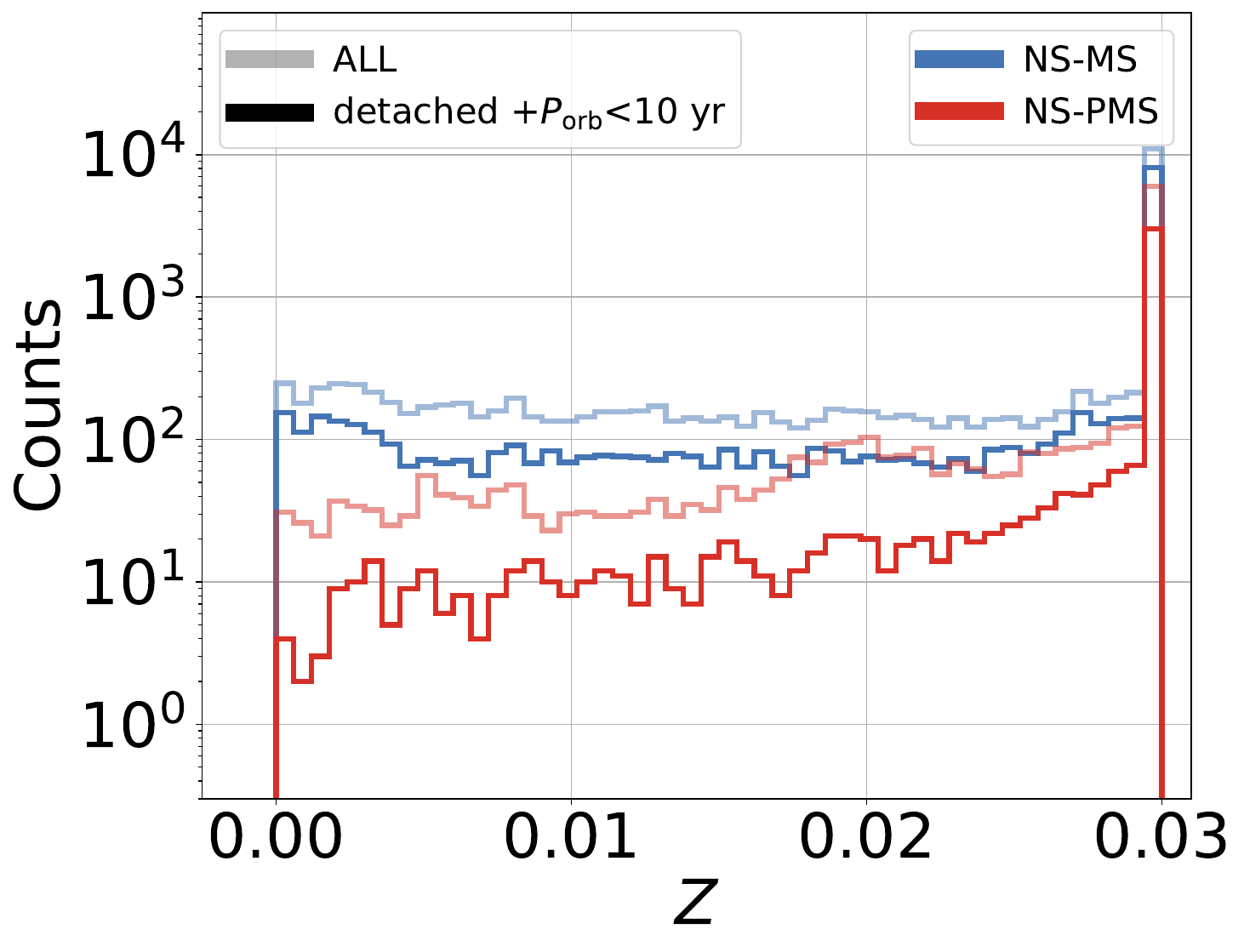}
    \caption{Same as \autoref{fig:intrinsic-met} but for the \delayed\ SN model.}
    \label{fig:intrinsic-met-delayed}
\end{figure}

\begin{figure*}[h!]
    \plotone{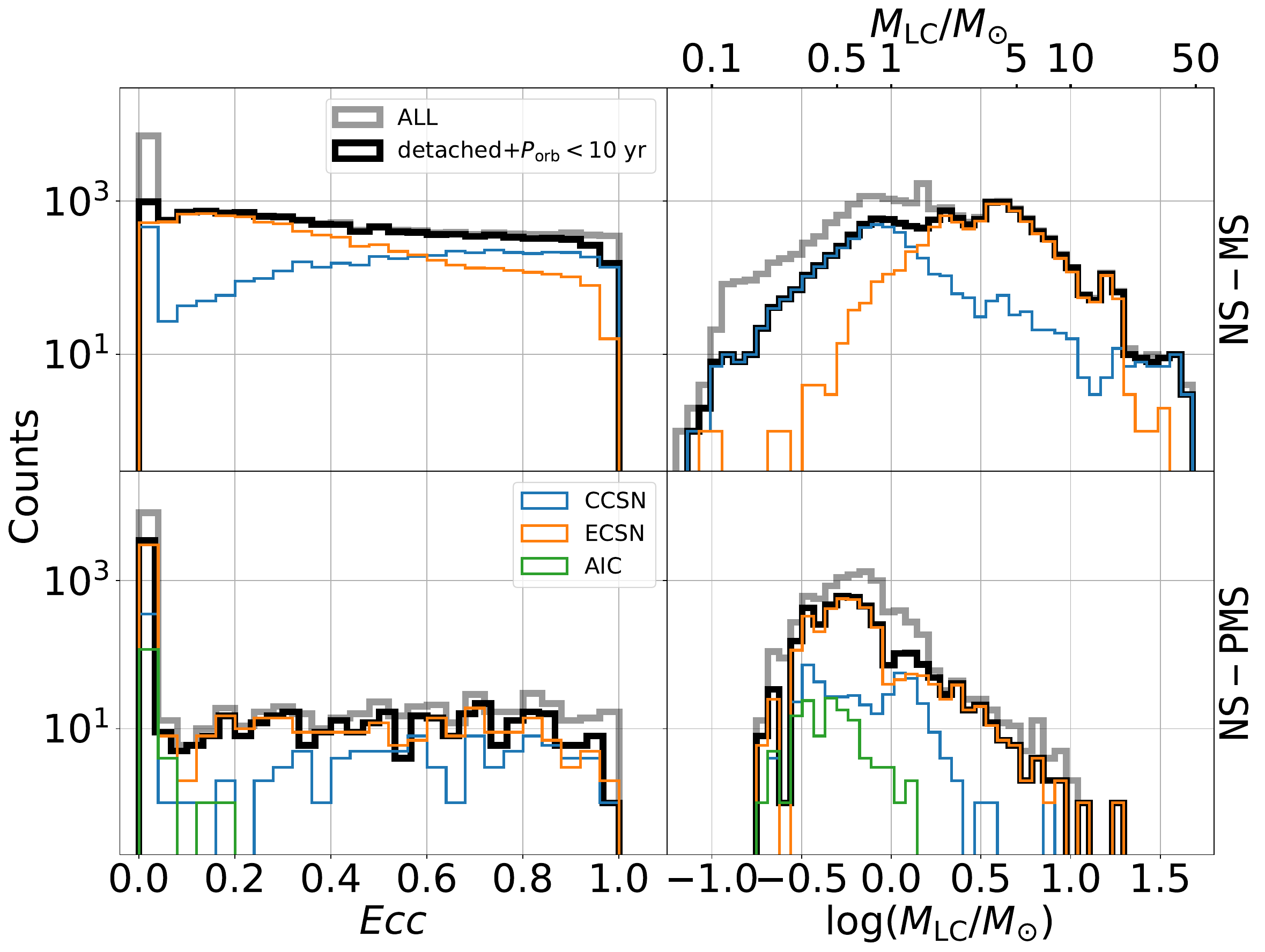}
    \caption{Same as \autoref{fig:intrinsic-ecc-mlc} but for the \delayed\ SN model.}
    \label{fig:intrinsic-ecc-mlc-delayed}
\end{figure*}

\begin{figure*}[h!]
    \plotone{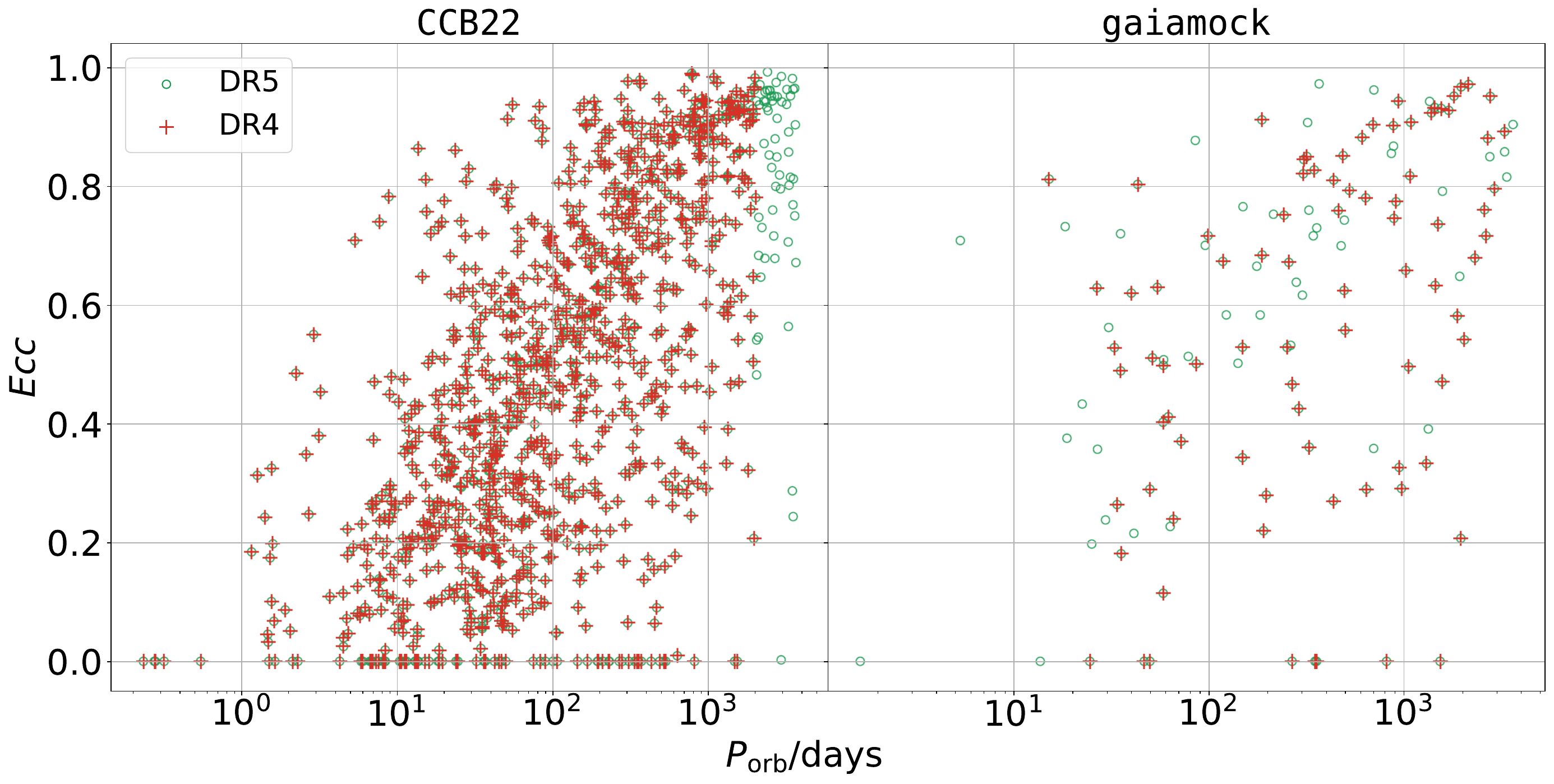}
    \caption{Same as \autoref{fig:porb-ecc-rapid} but for the \delayed\ SN model.}
    \label{fig:porb-ecc-delayed}
\end{figure*}

\begin{figure}[h!]
    \plotone{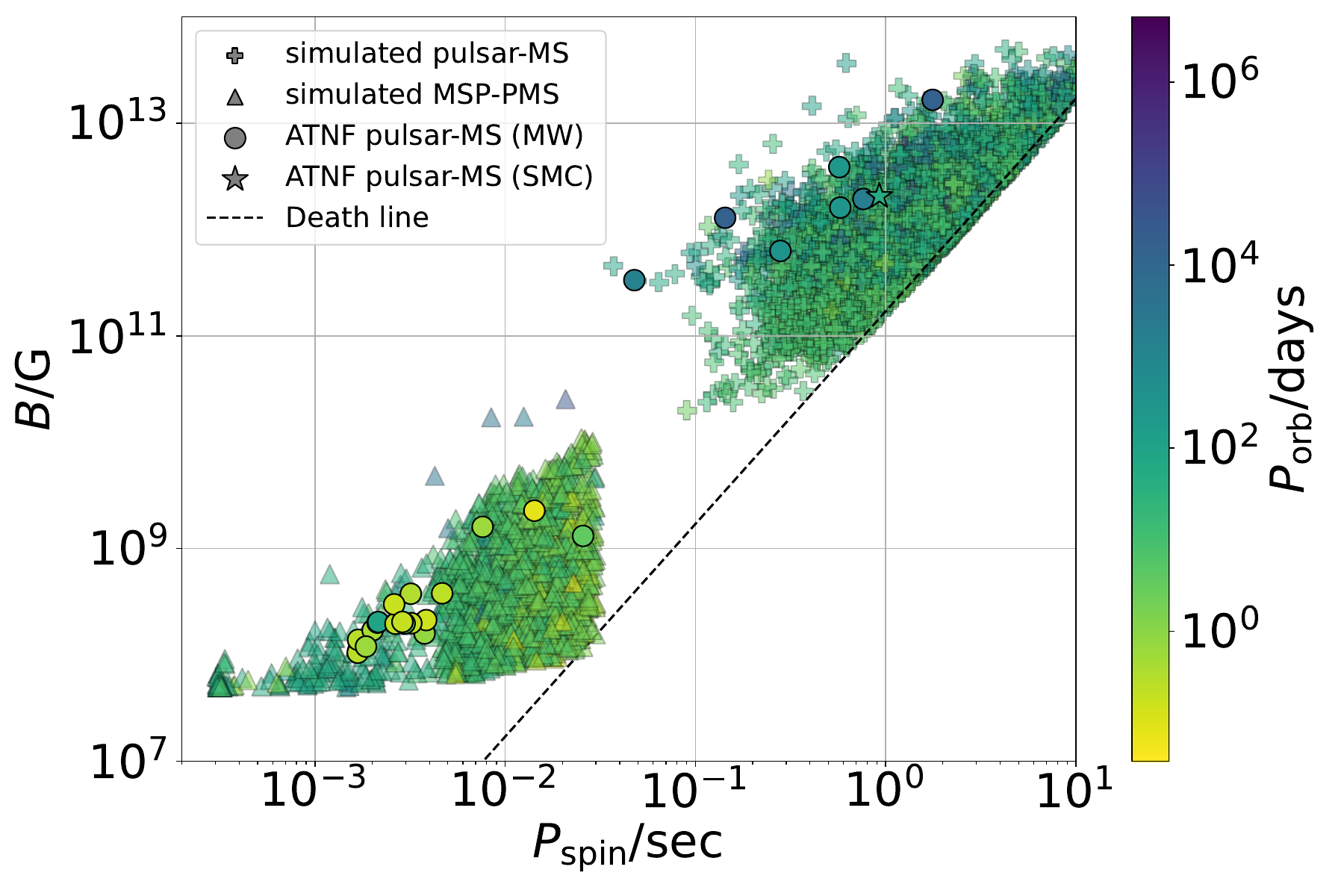}
    \caption{Same as \autoref{fig:atnf-B-P} but for the \delayed\ SN model.}
    \label{fig:atnf-B-P-delayed}
\end{figure}

\end{document}